\RequirePackage{fix-cm}
\documentclass[smallextended]{svjour3}       
\smartqed  
\usepackage{graphicx}
\usepackage[lined, linesnumbered, ruled]{algorithm2e}
\usepackage[cmex10]{amsmath}
\usepackage{amssymb}
\usepackage{subfigure}
\usepackage{multirow}
\usepackage{epstopdf}
\usepackage{wrapfig}
\usepackage{morefloats}
\usepackage[acronym,nonumberlist]{glossaries}
\newacronym{ACK}{ACK}{Acknowledgement}
\newacronym{ACU}{ACU}{Admission Control Unit}
\newacronym{AIFS}{AIFS}{Arbitration Inter Frame Space}
\newacronym{AP}{AP}{Access Point}
\newacronym{CBR}{CBR}{Constant Bit Rate}
\newacronym{CF}{CF}{Contention Free}
\newacronym{CFP}{CFP}{Contention Free Period}
\newacronym{CF-Poll}{CF-Poll}{Contention Free}
\newacronym{CP}{CP}{Contention Period}
\newacronym{CSMA}{CSMA}{Carrier Sense Multiple Access}
\newacronym{CW}{CW}{Contention Window}
\newacronym{CWmax}{CWmax}{Contention Window Maximum}
\newacronym{CWmin}{CWmin}{Contention Window Minimum}
\newacronym{DCF}{DCF}{Distributed Coordination Function}
\newacronym{EDCA}{EDCA}{Enhanced Distributed Channel Access}
\newacronym{EDCF}{EDCF}{Enhanced DCF}
\newacronym{FTP}{FTP}{File Transfer Protocol}
\newacronym{GI}{GI}{Generation Interval}
\newacronym{HC}{HC}{Hybrid Coordinator}
\newacronym{HCCA}{HCCA}{HCF Controlled Channel Access}
\newacronym{HCF}{HCF}{Hybrid Coordination Function}
\newacronym{HTTP}{HTTP}{Hypertext Transfer Protocol}
\newacronym{IEEE}{IEEE}{Institute of Electrical and Electronics Engineers}
\newacronym{MAC}{MAC}{Medium Access Control}
\newacronym{MSDU}{MSDU}{MAC Service Data Unit}
\newacronym{NAV}{NAV}{Network Allocation Vector}
\newacronym{PC}{PC}{Point Coordinator}
\newacronym{PCF}{PCF}{Point Coordination Function}
\newacronym{PF}{PF}{Persistence Factor}
\newacronym{PHY}{PHY}{Physical Layer}
\newacronym{PIFS}{PIFS}{PCF Inter Frame Space}
\newacronym{QAP}{QAP}{QoS-enabled Acces Point}
\newacronym{QBSS}{QBSS}{QoS-supporting Basic Service Set}
\newacronym{QoS}{QoS}{Quality of Service}
\newacronym{QSTA}{QSTA}{QoS-enabled Station}
\newacronym{RTS/CTS}{RTS/CTS}{Request to Send/Clear to Send}
\newacronym{SIFS}{SIFS}{Short Inter Frame Space}
\newacronym{SMTP}{SMTP}{Simple Mail Transfer Protocol}
\newacronym{TBTT}{TBTT}{Target Beacon Transmission Time}
\newacronym{TC}{TC}{Traffic Category}
\newacronym{TXOP}{TXOP}{Transmission Opportunity}
\newacronym{VBR}{VBR}{Variable Bit Rate}
\newacronym{WLAN}{WLAN}{Wireless Local Area Network}
\newacronym{TGe}{TGe}{IEEE 802.11 Task Group E}
\newacronym{UP}{UP}{User Priorities}
\newacronym{WM}{WM}{Wireless Medium}
\newacronym{CAP}{CAP}{Controlled Access Phase}
\newacronym{AC}{AC}{Access Categorie}
\newacronym{UGC}{UGC}{User-Generated Content}
\newacronym{TS}{TS}{Traffic Stream}
\newacronym{ATXOP}{ATXOP}{Adaptive TXOP Scheme}
\newacronym{AMTXOP}{AMTXOP}{Adaptive Multipolling TXOP Scheme}
\newacronym{TSPEC}{TSPEC}{TS Specification}
\newacronym{SI}{SI}{Service Interval}
\newacronym{EDD}{EDD}{Earliest-Due-Date}
\newacronym{RTP}{RTP}{Real-time Transport Protocol}
\newacronym{RTCP}{RTCP}{RTP Control Protocol}
\newacronym{OSI}{OSI}{Open Systems Interconnection}
\newacronym{AMC}{AMC}{Adaptive Modulation and Coding}
\newacronym{NPHCCA}{NPHCCA}{Non-Polling based HCCA}
\newacronym{LAN}{LAN}{Local Area Network}
\newacronym{RF}{RF}{Radio Frequency}
\newacronym{IR}{IR}{Infrared}
\newacronym{OFDM}{OFDM}{Orthogonal Frequency Division Multiplexing}
\newacronym{DSSS}{DSSS}{Direct Sequence Spread Spectrum}
\newacronym{HRDSSS}{HR/DSSS}{High-Rate DSSS}
\newacronym{STA}{STA}{Station}
\newacronym{IP}{IP}{Internet Protocol}
\newacronym{BSS}{BSS}{The Basic Service Set}
\newacronym{DEB}{DEB}{Deterministic Backoff}
\newacronym{QS}{QS}{Queue Size}
\newacronym{MSI}{MSI}{Maximum Service Interval}
\newacronym{SMA}{SMA}{Simple Moving Average}
\newacronym{TCP}{TCP}{Transmission Control Protocol}
\newacronym{UDP}{UDP}{User Datagram Protocol}
\newacronym{AID}{AID}{Associate Identifier}
\newacronym{PLCP}{PLCP}{Physical Layer Convergence Procedure}
\newacronym{PER}{PER}{Packet Error Rate}
\newacronym{FFBI}{FFBI}{Feed-Forward Bandwidth Indication}
\newacronym{F-Poll}{F-Poll}{Feasible Polling Scheme}
\newacronym{ARROW}{ARROW}{Adaptive Resource Reservation Over WLANs}
\newacronym{IFS}{IFS}{Interframe Space}
\makeglossaries
\begin{document}

\title{Feasible HCCA Polling Mechanism for Video Transmission in IEEE 802.11e WLANs
}


\author{Mohammed A. Al-Maqri  \and
Mohamed Othman \and
Borhanuddin Mohd Ali \and
Zurina Mohd Hanapi
}

\institute{Mohammed A. Al-Maqri . Mohamed Othman . Zurina Mohd Hanapi\at
Department of Communication Technology and Network, Universiti Putra Malaysia, Malaysia \\
mohdalmoqry@gmail.com (M.A. Al-Maqri), \\ mothman@upm.edu.my (M. Othman) 
\and
Borhanuddin Mohd Ali \at
Department of Computer and Communication Systems Engineering, Faculty of Engineering, Universiti Putra Malaysia, Serdang, 43400, Malaysia\\
}

\date{Received: date / Accepted: date}
\maketitle
\begin{abstract}
IEEE 802.11e standard defines two \gls{MAC} functions to support \gls{QoS} for wireless local area networks: \gls{EDCA} and \gls{HCCA}. \gls{EDCA} provides fair prioritized \gls{QoS} support while \gls{HCCA} guarantees parameterized \gls{QoS} for the traffics with rigid \gls{QoS} requirements. The latter shows higher \gls{QoS} provisioning with \gls{CBR} traffics. However, it does not efficiently cope with the fluctuation of the \gls{VBR} video streams since its reference scheduler generates a schedule based on the mean characteristics of the traffic. Scheduling based on theses characteristics is not always accurate as these traffics show high irregularity over the time. In this paper, we propose an enhancement on the \gls{HCCA} polling mechanism to address the problem of scheduling pre-recorded \gls{VBR} video streams. Our approach enhances the polling mechanism by feed-backing the arrival time of the subsequent video frame of the uplink traffic obtained through cross-layering approach. Simulation experiments have been conducted on several publicly available video traces in order to show the efficiency of our mechanism. The simulation results reveal the efficiency of the proposed mechanism in providing less delay and high throughput with conserving medium channel through minimizing the number of Null-Frames caused by wasted polls.
\end{abstract}
\section{Introduction}
\label{intro}
Recently IEEE 802.11 has become one of the massively deployed technology in the residential and public places such as apartments, stock markets, campuses, airports, etc. Due to some of its key features like deployment flexibility, infrastructure simplicity and cost effectiveness, there has been a recent trend toward providing an ubiquitous wireless access environment. This tendency leads to the presence of many multimedia applications with various traffic characteristics. In the future, it is widely expected that next generation wireless networks will be carrying a large portion of encoded video streams, two-third of all traffics in the networks will be video by 2017 according to Cisco Visual Networking Index \cite{indexglobal}. IEEE 802.11 WLANs \cite{IEEEStand1999} were designed for the transmission of the best effort services which are no longer sufficient to meet the vast growth of time-bounded services that require rigorous \gls{QoS} requirements such as channel bandwidth, delay and jitter \cite{Aljubari2013,Saif2013}. Since \gls{MAC} layer functions are not QoS-oriented, guaranteeing \gls{QoS} in a such layer has become a challenging task. \gls{TGe} has presented IEEE 802.11e standard to improve the \gls{QoS} support of multimedia streaming over WLANs.

The IEEE 802.11e introduced differentiated \gls{QoS} services through a novel \gls{HCF} which is included into the recent standard, released on 2007 \cite{IEEEStandard2007}. A new revised version with technical enhancements on \gls{MAC} and Physical layer has been launched on 2012 \cite{IEEEStandard2012}. The \gls{HCF} promotes the channel access modes of IEEE 802.11, \gls{DCF} and \gls{PCF} into \gls{EDCA} and \gls{HCCA} respectively. \gls{EDCA} provides prioritized \gls{QoS} support by defining multiple \glspl{AC} with AC-specific \gls{CW} and \gls{AIFS} to identify different levels of priority. \gls{QoS} is achieved through classifying delay-sensitive application such as video and voice into the highest priority ACs so they experience smaller backoff times. However, \gls{EDCA} mechanism is unsatisfactory for supporting real-time applications with rough \gls{QoS} requirements, especially in heavily loaded networks, where collision possibility is high. Recently, a number of efforts have been presented to discuss the deficiency of supporting \gls{QoS} in \gls{EDCA} mode such as \cite{Andreadis2012,Shakir2013,Pang2013}.

In \gls{HCCA}, parameterized \gls{QoS} support is achieved through scheduling \glspl{QSTA} traffics in a \gls{BSS} based on their negotiated \gls{QoS} with the \gls{HC} which is usually collocated with the \gls{AP}. Newly joined \glspl{QSTA} are admitted to the system, asserting that previously admitted services are not jeopardized. \gls{HCCA} is promising scheme for supporting \gls{QoS} for delay-constrained applications such as VoIP and video streams compared to its counterpart (\gls{EDCA}). This is due to the fact of eliminating the backoff counter overhead and the collision caused by the hidden node which is inherent in distributed access mode.

Although, the reference \gls{HCCA} schedules traffics upon their negotiated \gls{QoS} requirements in the first place, it is only efficient for \gls{CBR} applications such as \gls{CBR} G.711 \cite{G7111988} audio streams and H.261 \cite{MPEG11997} video (MPEG-1). However, it is not convenient to deal with the fluctuation of the \gls{VBR} traffic such as H.263 \cite{H2631996} video streams and G.718 \cite{G7182009} audio traffic, where neither the packet size nor the packet generation time is constant. This consequently leads to a remarkable increase in the end to end delay of the delivered traffics and degradation in the channel bandwidth utilization as well.

The \gls{HC} which resides in \gls{QAP} maintains separate queues for the downlink traffic streams while the uplink streams are maintained in \glspl{QSTA}' queues. For this reason, the \gls{HC} can allocate time resources for its queues easily, yet it is unable to predict the amount of the \gls{VBR} uplink traffics due to the fact that it is physically separated from the \glspl{QSTA}. Several mechanisms such as \cite{Lee2009,Jansang2011,Cecchetti2012} have been recently proposed to remedy the deficiency of the \gls{HCCA} reference scheduler in supporting \gls{QoS} for \gls{VBR} video traffics. However, these enhancements still not sufficient to cope with the fast fluctuating nature of high compressed video applications due to the difficulty of accurately predict the \gls{VBR} traffic profile. Recently, \cite{Maqri2015} presents a multi-polling approach to enhance the QoS provision of the VBR videos that show variability in packet size with fixed packet inter-arrival time such as MPEG-4.

With the increase of Internet web applications in the wireless mobile devices, the \gls{UGC} such as pre-recorded video streams have become more prominent nowadays. To the best of our knowledge, the scheduling of uplink pre-recorded continuous media in \gls{HCCA} has not been addressed efficiently despite the fast growth of uplink streams of the \gls{UGC} on the Internet such as pre-recorded video streams. In this paper, we present an enhancement on the \gls{HCCA} polling mechanism. The proposed mechanism adjusts the legacy polling based on the feedback information sent to the \gls{HC} in order to accommodate to the fast changing of the \gls{VBR} traffics which show variability in packet generation interval such as H.263 streams. This mechanism makes use of the queue size field of \gls{QoS} data frame in the \gls{MAC} header of the IEEE 802.11e to carry this information to the \gls{HC}, this is discussed in details in Section \ref{sec:Fpoll}.

The rest of this paper is organized as follows:
Section ~\ref{sec:RelatedWork} explains the reference \gls{HCCA} mechanism and demonstrates its deficiency in supporting \gls{VBR} Video streams and discusses some enhancements on its polling mechanism. Section ~\ref{sec:Fpoll} explains the proposed algorithm. The performance evaluation and discussion is presented in Section ~\ref{sec:perfEval}. Section ~\ref{sec:conclusion} concludes the study presented in this paper.
\section{Background and Related Work}
\label{sec:RelatedWork}
In this section, the reference scheme of the IEEE 802.11e \gls{HCCA} along with some characteristics of the \gls{VBR} video application are reviewed. The deficiencies of \gls{HCCA} in supporting \gls{VBR} and some related work in enhancing its performance are also discussed.
\subsection{IEEE 802.11e HCCA Mechanism}
\begin{figure*}[!h]
\centering
\includegraphics[width=\linewidth]{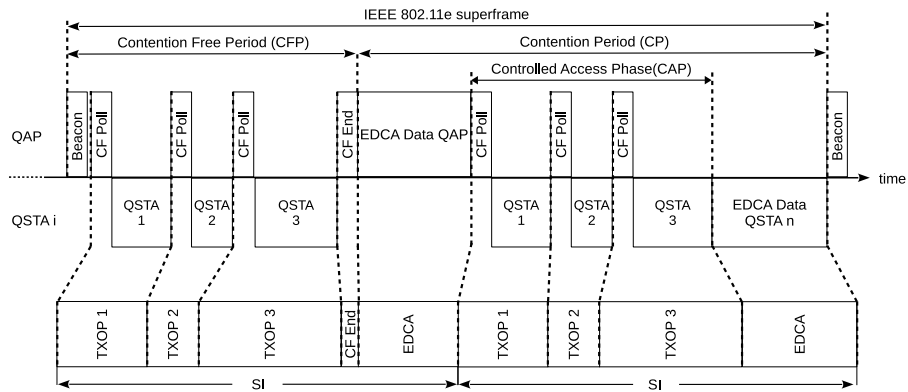}
\caption{Example of an 802.11e Superframe, Optional Contention-free Period and Contention Period}
\label{fig001}
\end{figure*}

In IEEE 802.11e, a beacon is transmitted periodically to all stations in the \gls{BSS} for synchronization purpose. The time between two subsequent beacons forms a so-called superframe. Service delivery occurs during the superframe in two periods, \gls{CFP} and \gls{CP}. A station shall transmit its traffics within a duration of time called \gls{TXOP} which is a time duration reserved for a \gls{QSTA} to deliver its \glspl{MSDU}. The \gls{TXOP} obtained via contention-based access referred to EDCA-TXOP, while in controlled medium access the \gls{TXOP} granted to the \gls{QSTA} by the \gls{HC} so that it called polled \gls{TXOP}. \mbox{Figure ~\ref{fig001}} illustrates an example of the alternation of one  controlled medium access followed by a contention-based period with one \gls{QAP} and three \glspl{QSTA}. 
Note that the controlled medium access not only occur during the \gls{CFP} but also during the \gls{CP} when the medium is sensed idle for a period of \gls{PIFS}. For this reason, supporting \gls{QoS} in \gls{HCCA}  mode is improved compared to its ancestor, \gls{PCF}, where the controlled transmission only occurs during \gls{CFP}. The station in \gls{HCCA} mode can only transmit its data packets for a duration declared in the poll frame.
\subsection{Reference Design of HCCA}
When a \gls{QSTA} desires to deliver its real-time \gls{TS} during the contention free phase it sends an ADDTS-Request to the \gls{QAP} declaring its desired \gls{QoS} requirements for the particular \gls{TS}. The required QoS parameters are carried in the traffic specification \gls{TSPEC} fields. Accordingly, \gls{QAP} will try to satisfy these requirements with an attempt to conserve the \gls{QoS} of the already admitted flows. The accepted ADDTS-Request will be replied by an ADDTS-Response and the particular station will be added to the polling list in the \gls{QAP}. A list of the mandatory \gls{TSPEC} parameters is presented in Table \ref{tab:TSPECsymbol}.
\begin{table}[!h]
\caption {\gls{TSPEC} and Scheduling Parameters Symbols}
\centering
\begin{tabular}{lp{2cm}l}
	\hline
	Symbol 	& Unit		& Description \\ \hline
	$\rho$	& bit/sec	& Mean Data Rate\\
	$L$ 	& bytes		& Nominal MSDU Size \\
	$M$		& bytes		& Maximum MSDU Size \\
	$D$ 	& sec		& Delay Bound\\
	$SI$ 	& sec		& Service Interval\\
	$mSI$ 	& sec		& Minimum Service Interval\\
	$MSI$ 	& sec		& Maximum Service Interval\\
	$R$ 	& bit/sec	& Physical Transmission Rate\\
	$BI$	& sec		& Beacon Interval \\  
	$O$		& sec		& PHY and MAC Overhead \\
	$N$		& Number 	& Number of packets \\
	$T$		& sec		& Superframe duration \\ 
	$T_{CP}$& sec		& Contention-based duration \\ \hline 
	
\end{tabular}
\label{tab:TSPECsymbol}
\end{table}

%
%
%
%
%
%
%

Upon receiving an ADDTS-Request from a \gls{QSTA}, the \gls{HCCA} reference scheduler goes through the following steps:
\begin{enumerate}
\item \textit{SI Assignment}

\label{SIassign}
In the \gls{HCCA} reference scheduler, \gls{SI} is computed as a submultiple of the beacon interval $BI$ which is the minimum of the maximum SIs for all previously admitted traffics including the incoming traffic as shown in Equation (\ref{si})
\begin{equation}
\label{si}
SI = \frac{BI}{\left \lceil\frac{BI}{MSI_{min}} \right \rceil}
\end{equation}
where $MSI_{min}$ is calculated in Equation (\ref{minSI})	
\begin{equation}
MSI_{min} = min(MSI_{i}) , i \in [1, n]
\label{minSI}
\end{equation}	
and $n$ is the number of admitted \glspl{QSTA}' traffic streams and $MSI_{i}$ is the maximum $SI$ of the $i^{th}$ stream.
\\

\item \textit{TXOP Allocation} \\
\gls{HC} allocates different \gls{TXOP} to each admitted station so as to transmit their data with respect to the declared \gls{TSPEC} parameters. This \gls{TXOP} is calculated for each station as follows:

Firstly, for the $i^{th}$ station, the scheduler calculates the number of MSDUs that may arrive at $\rho_{i}$ as in Equation (\ref{n}) :
\begin{equation}
N_{i}=\left \lceil \frac{SI\times\rho_{i}}{L_{i}} \right \rceil,
\label{n}
\end{equation}
where $L_{i}$ is the nominal \gls{MSDU} for the $i^{th}$ station.	
Then the \gls{TXOP} duration of the particular station, $TXOP_{i}$, is calculated as the maximum of the time required to transmit a number $N_{i}$ of \glspl{MSDU} or time to transmit one maximum \gls{MSDU} at a physical rate $R_{i}$, as stated in Equation \eqref{txop}:
\begin{equation}
TXOP_{i}=max\left (\frac{N_{i} \times L_{i}}{R_{i}} + O, \frac{M}{R_{i}} + O \right)
\label{txop}
\end{equation}
where $O$ is the overhead, including \gls{MAC} and physical headers, \glspl{IFS}, the acknowledgment and poll frames overheads.
\\
\item \textit{Admission Control} \\
\gls{ACU} manages the admission of the \glspl{TS} with insuring that the \gls{QoS} of the already admitted TSs is maintained. When a new \gls{TS} demands an admission, the \gls{ACU} : 1) obtains a new $SI$ as shown in step \ref{SIassign} and calculates the number of MSDUs expected to arrive at the new $SI$ using Equation (\ref{n}). 2) the \gls{ACU} then calculates the $TXOP_{i}$ for the particular \gls{TS} as in  Equation (\ref{txop}). Finally, the \gls{ACU} admits the \gls{TS} only if the following inequality satisfied:
\begin{equation}
\label{eq:ACU}
\frac{TXOP_{n+1}}{SI}+\sum_{i=1}^{n} \frac{TXOP_{i}}{SI}\leq \frac{T- T_{CP}}{T}
\end{equation}
where $n$ is the number of admitted traffic streams, $n+1$ is the index of the incoming \gls{TS}, $T$ is the beacon interval and $T_{CP}$ is the duration reserved for \gls{EDCA}, contention-period. 

The \gls{HC} sends an acceptance message, ADDTS-Response, to the requested \gls{QSTA} if the Equation \eqref{eq:ACU} is true or send a rejection message otherwise. The accepted \gls{TS} will be added to the polling list of the \gls{HC}.
\end{enumerate}
\subsection{VBR Video Traffic Application}
\label{sec:VBRchar}
Video compression has been widely used in today's wireless system transmissions due to the limitation of transmission rate in the wireless networks. Several encoding techniques have been adopted to produce compressed video with different quality levels like Motion Picture Experts Group type 1 (MPEG-1) and MPEG-2. A  high picture quality can be obtained by encoding video at several megabits per second bit rate, yet it is not appropriate for transmitting over the limited wireless medium \cite{H2631996}. Thus, encoding schemes providing lower bit rate has been emerged to produce video with an acceptable picture quality at even lower than 64 kpbs bit rate. 

H.263, is an example of low rate, video compression, achieves more than 50\% reduction in the bit rate required to represent the equivalent video quality compared to H.261. H.263 encodes the video sequence at a reference frame rate, e.g. 25 frames/Sec, i.e. every \mbox{40 ms}. However, in order to meet the smaller bit rate, H.263 encoder skips some frames which consequently results in a variable frame interval. A fragment of 10 frames of H.263 trace file of \textit{Silence of the Lambs} film exhibits the fact of frame skipping is shown in Table \ref{tab:traceFrag}. The number of the skipped frames tends to be higher as the target bit rate goes smaller, which means the higher varying feature of video traffic. Such kind of video streaming served poorly by \gls{HCCA} based on fixed \gls{TSPEC} parameters negotiated at the traffic setup. The following section discusses the issue of overpolling the stations in the reference design of the \gls{HCCA}.
\begin{table}[!h]
\centering
\caption {A Fragment of Silence of the Lambs Trace File Encoded Using H.263 at 256kbps Target Bit Rate \cite{Fitzek2001}}
\begin{tabular}{ccc}
	\hline
	Frame period (ms)& Frame type & Frame size (bits) \\ \hline
	0    & I  & 12539 \\
	360  & P  & 3981  \\
	600  & PB & 6203  \\
	760  & PB & 5884  \\
	1000 & PB & 6749  \\
	1160 & PB & 6425  \\
	1400 & PB & 7849  \\
	1640 & PB & 5983  \\
	1800 & PB & 6183  \\
	2040 & PB & 7052  \\ \hline
\end{tabular}
\label{tab:traceFrag}
\end{table}
\subsection{Preliminary Study of \gls{HCCA} Polling Mechanism}
\label{sec:preStudy}
Since \gls{HCCA} schedules \gls{QSTA} based on negotiated \glspl{TSPEC} which represent their mean traffic characteristics, it is not efficient to cope with the variable profile of \gls{VBR} traffic streams. Polling all \glspl{QSTA} at the same \gls{SI} period may cause degradation in the channel utilization since some \glspl{QSTA} are not ready to send data and thus reply to poll by Null-Frames. Consider Figure \ref{fig:wastingPolls} where number of \gls{VBR} traffics are scheduled in one \gls{SI}. The beacon interval is 200 ms and three \glspl{QSTA} are scheduled every 40 ms. Assume that the first \gls{SI} begins at time 0 and all \glspl{QSTA} commence their traffics at that time. In this example, all \glspl{QSTA} will use their \gls{TXOP} duration at the first \gls{SI} since they start their traffic and must have packets to send. However and due to the varying feature of the \gls{VBR} traffics, at some SIs there might be one or more \glspl{QSTA} with no packet to send thus they are considered as over-polled. Consider the second \gls{SI} at time 40 ms $QSTA_{1}$ has no data which will reply by Null-frame. In the third \gls{SI} it is even worse where only $QSTA_{3}$ utilizes the poll and transmit data packets while $QSTA_{1}$ and $QSTA_{2}$ will reply by Null-frames and so on. Polling a \gls{QSTA} with no data will remarkably increase the poll overhead which consists of transmitting one poll frame, a Null-frame and an \gls{ACK}.

\begin{figure}[!h]
	\centering
	\includegraphics[scale=0.5]{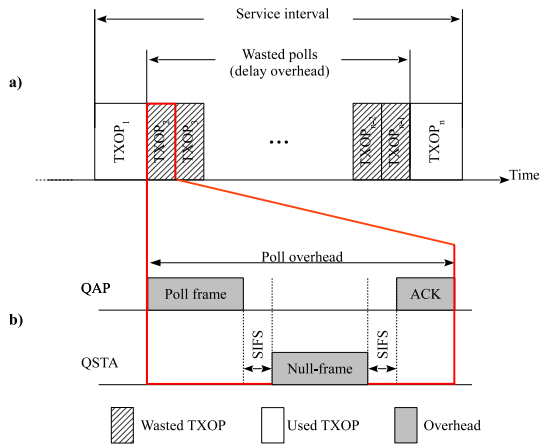}
	\caption{Unwanted Delay Caused by Wasted Polls by Unready QSTAs with VBR Traffics in The Reference Design of HCCA.}
	\label{fig:pollOH}
\end{figure}

\begin{figure*}[hbtp]
\centering
\includegraphics[width=\linewidth]{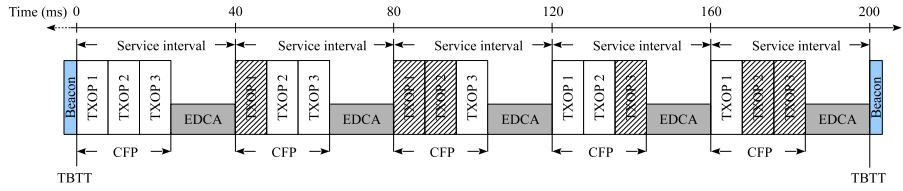}
\caption{Wasting Polls in \gls{VBR} Traffics, Example of Three \glspl{QSTA} Each of Which Has one \gls{VBR} Traffic}
\label{fig:wastingPolls}
\end{figure*}

Consider the case illustrated in Figure~\ref{fig:pollOH} where there are $n$ number of QSTAs are polled in the current SI in order. Assume that only $QSTA_{1}$ and $QSTA_{n}$ have packet in their transmission queue and the rest of the stations have no packets. In this case, $QSTA_{n}$ will have to wait for all stations ahead to transmit their Null-frames. This inherent issue in HCCA occurs because QAP is unaware about the current change in VBR traffic profile.
It worth noting that inimizing the number of wasted polls may reduce the delay in both \gls{EDCA} and \gls{HCCA} and boosts the channel utilization as well. For instance, at the third \gls{SI} preventing polling $QSTA_{1}$ and $QSTA_{2}$ will lead to poll $QSTA_{3}$ earlier and thus reduce the end-to-end delay. On the other hand, at time 160 ms the wasted time of polling $QSTA_{2}$ and $QSTA_{3}$ can be transfered to \gls{EDCA} period, which enhance the system channel utilization and the packet delay.
The \gls{QAP} over-polls \gls{QSTA} because of lack of information about the backlogged packets on \glspl{QSTA} queues. This is normal in a controlled-access mode where the station only transmits upon receiving a poll from \gls{QAP}. Many approaches have been conducted to alleviate the overhead caused by sending polls to unready stations with \gls{VBR} traffics we present some of them in the following section.
\subsection{Enhancements on \gls{HCCA} Scheduler for Supporting \gls{VBR} Traffics}
It is hard for the \gls{HC} to accurately schedule the uplink TSs for the reason that \gls{HC} is physically separated from the \glspl{QSTA} and because of the stochastic behavior of the \gls{VBR} traffics. Several approaches proposed in the literature to assist the \gls{HCCA} scheduler accommodate with the \gls{VBR} traffic fluctuation over the time. In this section, we discuss some of the enhancements on the \gls{SI} assignment and the polling mechanism of \gls{VBR} traffics in the reference design.

Scheduling Based on Estimated Transmission Times-Earliest Due Date (SETT-EDD) \cite{Grilo2003} is deemed one of the exemplary solution proposed in the literature. SETT-EDD schedules the TSs in the system considering the delay of the head-of-line packets of the scheduled TSs by the mean of using the well known algorithm in multimedia, Delay Earliest Due Date (EDD). In the case of the downlink traffics, the \gls{HC} directly obtains the delay of the TSs head-of-line packet delay since it maintains these traffics at its transmission queues. In the case of the uplink traffics, the SETT-EDD is unable to read these values, thus it estimates them using a token-bucket algorithm affiliated with each \gls{TS}. 

This scheme succeeds in reducing the packet delay and the packet loss. Notwithstanding in high bursty traffics the average \gls{TXOP} is not efficient and may lead to high transmission delay, since it considers only average \gls{TXOP} transmission. Besides, inaccurate estimation of network status may lead to degradation in bandwidth utilization.

\gls{ARROW} scheme \cite{Skyrianoglou2006} has received considerable attention in the literature since it considers the actual packets buffered at stations rather than estimate the buffered data. \gls{ARROW} utilizes the \gls{QS} by \gls{QSTA}, which is part of \gls{QoS} Data frames introduced by 802.11e standard \cite{IEEEStandard2007}. Thus to inform the \gls{HC} about backlogged packets in the their transmission queue at the previous \gls{SI}. A scheduler exploits this information and allocates TXOPs to \glspl{QSTA} in such a way that attempt to clear their queues as long as they comply with the declared \gls{TSPEC}. \gls{ARROW} achieves more bandwidth utilization and higher throughput compared to SETT-EDD. However, calculation of SIs based on the mean parameters declared in \gls{TSPEC} might cause high packet loss when high variable packet interval is used, such as low bit rate H.263 video streams \cite{Huang2009}.

A four-way-polling \cite{Huang2010} is a feedback approach in which information about the backlogged information is inquired by \gls{QAP} about the buffer occupancy of the uplink TSs. Unlike \gls{ARROW}, where the feedback information and the respective \gls{SI} assignment is done separately in two consequent \gls{SI}, in this approach the \gls{QAP} explicitly inquires the backlogged information through transmitting a request packet and the \gls{QSTA} reports its buffer in a response packet. Accordingly, \gls{QAP} sends an allocation packet with the \gls{TXOP} duration enough to serve the buffered load at that station. The Four-way-polling algorithm succeeds in minimizing the packet delay and loss. However, exchanging additional packets may raise the risk of being attacked and lead to wasting channel time, especially when the \gls{SI} is small, in such case the polling happened more frequently.

In \cite{Chen2011}, the authors proposed a modification on \gls{HCCA} named Non-polling based \gls{HCCA} \gls{NPHCCA} to alleviate the inherited issue on wasted polls on the reference design. The key idea is to let stations report their transmission requirement (buffered packets) during reserved channel called Req\_Time after \gls{QAP} broadcasts the beacon frame. Stations that succeed to send their request frames will be notified in the next beacon for transmission. \gls{NPHCCA} involves modifications in the reference design data structure by adding a new frame, called transmission request frame which conveys information about the station backlogged packets to the \gls{QAP}. In addition, in the \gls{QAP}, a new table is defined to maintain information about the station's transmission status to help \gls{QAP} decide an appropriate  scheduling sequence of stations with respect to their \gls{QoS} needs. Besides, additional field extends the beacon frame which is used by \gls{QAP} for transmission notification.  

Results reveal that \gls{NPHCCA} has reduced the access delay of the stations by avoiding polling stations that have no data. Average transmission delay is also minimized due to the fact that \gls{AP} assigns a real-time \gls{TXOP} to stations to transmit their frames with regards to their requests. However, the performance of \gls{HCCA} and \gls{NPHCCA} are almost equivalent when the network is heavily loaded. Accordingly, references \cite{Lo2003,Kim2005,Leu2005,Kim2006,Hsieh2009} tend to reduce the poll overhead by sending a multiple poll frame to all stations instead of periodically transmitting one poll to each.

The Enhanced \gls{EDD} \cite{Lee2009}. The \gls{TXOP} is dynamically allocated based on the data rate reported to \gls{QAP} in \gls{TSPEC} of each \gls{TS}. Information about the backlogged packets is delivered at the previous \gls{SI} just like \gls{ARROW}. The \gls{TXOP} of a station $i$ can be computed as in Equation~\eqref{eq:EddTXOP}. The scheduler also adaptively advances the start of the \gls{SI} when it determines that 1) there are backlogged packets at \gls{QSTA} transmission queues which implies that packets generated at the previous \gls{SI} have not served hence the new \gls{SI} will be advances by a time enough to clear the queue as in Equation \eqref{eq:mSI1}. 2) the allocated \gls{TXOP} is not fully utilized by the \gls{QSTA} in this case the \gls{SI} will be preponed by the residual time of the previous \gls{TXOP} as in Equation \eqref{eq:mSI2}.

Simulation results reveal that Enhanced \gls{EDD} scheduler has outperforms both \gls{HCCA} and SETT-EDD schedulers in terms of end-to-end delay. This is due to preventing backlogged packets from unnecessarily waiting caused by fixed mSI and \gls{MSI}.
\begin{equation}
TXOP_{i}=TXOP^i_{avg} + TD_{i}
\label{eq:EddTXOP}
\end{equation}
where $TXOP^i_{avg}$ is the \gls{TXOP} calculated for the station $i$ considering the data rate since the previous \gls{SI} until the current time whereas the $TD_{i}$ is the \gls{TXOP} required to transmit the backlogged packets of the \gls{TS}.
\begin{equation}
mSI^i_{new}=mSI^i - TD^i_{cur}
\label{eq:mSI1}
\end{equation}
where the $mSI^i_{new}$ is the new \gls{SI} for the station and $TD^i_{cur}$ is the \gls{TXOP} duration required to send the backlogged packets at station $i$.
\begin{equation}
mSI^i_{new}=mSI^i - TD^i_{free}
\label{eq:mSI2}
\end{equation}
where $TD^i_{free}$ is the unused \gls{TXOP} duration of at the previous \gls{SI}.
\section{Feasible Polling Scheme (F-Poll)}\label{sec:Fpoll}
In F-Poll, the exact arrival time of the next frame of the uplink stream is obtained at the \gls{MAC} layer of the station from its application layer through cross layering concept. For this reason we call it \gls{F-Poll}, in which accurate information about the next inter-arrival  time is transmitted to the \gls{QAP} for enhancing the scheduling of the \glspl{TS}. Upon the reception of the data frame a decision is made about either polling the respective station in the next \gls{SI} or not, to prevent polling stations that are not ready to transmit and consequently minimize the packet access delay and maximize the channel utilization.

\gls{F-Poll} enhances the polling scheme of the \gls{HCCA} reference design to accurately poll stations with encoded video streams. We present in this section the description of the scheduling process of the \gls{F-Poll} in both station and access point.

\IncMargin{1.2em}
\begin{algorithm}[!h]
\DontPrintSemicolon
\textbf{Input:}

$stations$, a list of $N$ stations in the polling list of the HC\;
$arrivals$, a list of next packet arrival time for each $station_{i}$ in $stations$ where $i=1..N$\;
WHEN receiving a data packet from a $station_{i}$

$arrival_{i} \leftarrow arrival_{i} - SI $

\For{each CAP} {
	WHEN no data packet received from $station_{i}$
	
	$arrival_{i} \leftarrow -1$
	
	$i \leftarrow 1$
	
	\While{$i \leq N$}{
		\uIf{first CAP of $station_{i} $}{
			Poll $station_{i}$\;
		}
		\uElseIf{$arrival_{i} \leq 0$}{
			Poll $station_{i}$\;
		}
		\Else{
			$arrival_{i} \leftarrow arrival_{i} - SI$\;
		}
		$i \leftarrow i + 1$
	}
}
\caption{F-Poll Scheme Pseudo Code}
\label{algo0105}
\end{algorithm}
\DecMargin{1.2em}
\subsection{Scheduling Actions at the Station}
\label{subsec:5_OpatSta}
At the station, information about the next frame arrival time is obtained from the application layer via cross-layering. The information of the next frame arrival time can be obtained based on the deployment at the application layer. For instance, in \gls{RTP} defined in \cite{RTP2003} which is suitable for applications transmitting real-time data each \gls{RTP} packet has a sequence number and timestamp. The timestamp represents the packet arrival time to be used in F-Poll scheme, as the whole or prefetched part of the video is known before the streaming commences. At the \gls{MAC} layer, the feedback information is carried in the \textit{Queue Size} (\textit{QS}) field introduced by IEEE 802.11 standard \cite{IEEEStandard2007} which is a part of the \gls{QoS} Control field of the \gls{QoS} data frame. The \textit{QS} field is exploited in this scheme for sending information about the next frame arrival time to the \gls{QAP} for scheduling purpose.
\subsection{Scheduling Actions at the Access Point}
\label{sec:5_OpatAP}
After the traffic setup, the \gls{QAP} sends the first poll frame granting the \gls{QSTA} a \gls{TXOP} duration and the station will accordingly transmit the first packet of its traffic to the \gls{QAP}. Note that the inter-arrival time between encoded video traffic frames is a multiple of a fixed interval (e.g. 40 ms) depends on the encoding parameters. That is to say, it is expected to receive only one packet at a multiple of the designated interval.

\begin{figure}[!h]
\centering
\includegraphics[width=\linewidth]{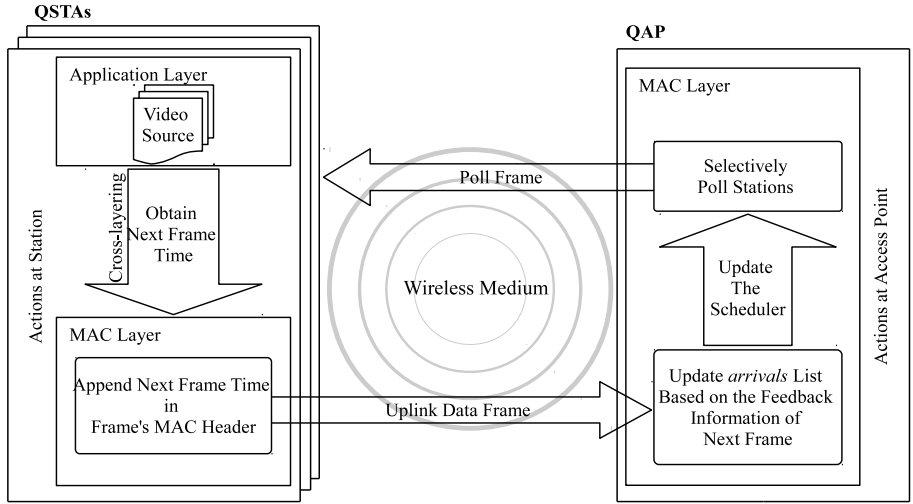}
\caption{F-Poll Scheme Framework}
\label{fig:FPollFramework}
\end{figure}

\begin{figure}[htpd]
	\centering
	\includegraphics[scale=.8,angle=90]{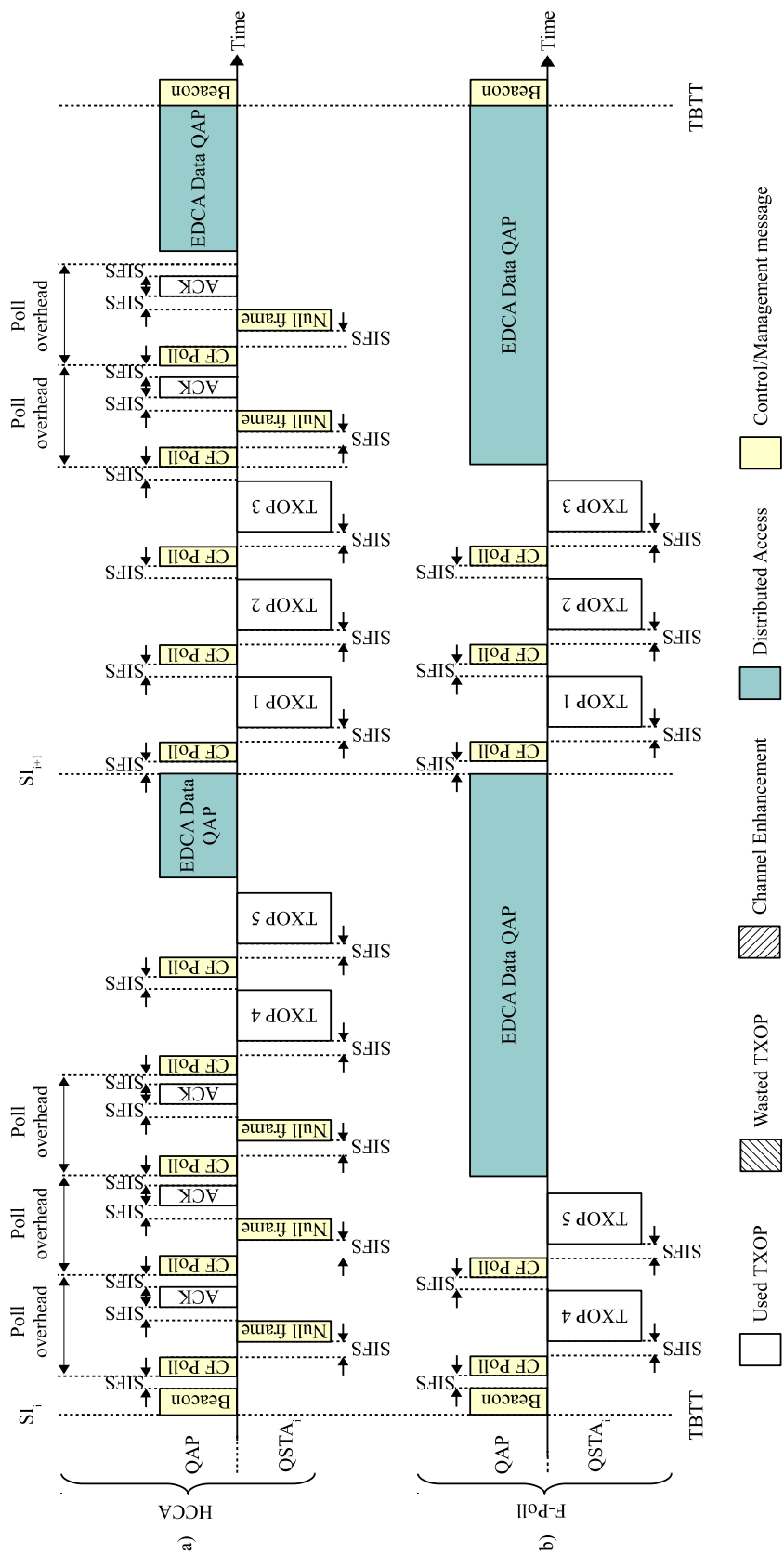}
	\caption{F-Poll Scheme}
	\label{fig:FPollMechanism}
\end{figure}

At the beginning of each \gls{CAP}, the \gls{QAP} goes through the polling list, which maintains a list of all admitted QSTAs in the system and behaves according to two cases. The first case is when a data packet received from the $QSTA_{i}$ in the previous CAP/SI period, the arrival time ($arrival_{i}$) of the next frame is obtained. Then, a value of \gls{SI} is deducted from it to compute the time left to the next frame at the $QSTA_{i}$. The station is polled only if the $arrival_{i}$ is less than or equal to zero which implies that the next frame is generated at $QSTA_{i}$ and is waiting for a poll to transmit. The second case is when no data packet received due to loss, the \gls{QAP} will keep polling the \gls{QSTA} at the original \gls{SI} rate regardless if it has data to transmit or not. The information about the arrival time of next frame held in the first received packet from the correspond \gls{QSTA} after loss will be ignored and \gls{F-Poll} scheme will resume its operation when the second packet after loss received. Algorithm~\ref{algo0105} reports the pseudo code of \gls{F-Poll} scheme and Figure~\ref{fig:FPollFramework} depicts F-poll framework. 
Figure~\ref{fig:FPollMechanism} illustrates the polling overhead introduced by polling QSTAs that have no data to transmit and how \gls{F-Poll} tackle this issue. 
The \gls{HCCA} Polling scheme in Figure~\ref{fig:FPollMechanism}(a) $SI_{i}$, polls QSTAs regardless whether they have date to transmit or not, thus $QSTA_{1}$, $QSTA_{2}$ and $QSTA_{3}$ will not use their polls and replied by a null frame instead. Likewise, in $SI_{i+1}$, the stations $QSTA_{1}$, $QSTA_{2}$ and $QSTA_{3}$ used their where $QSTA_{4}$, $QSTA_{5}$ will waste the channel time sending null frames.

\begin{figure}[!h]
	\centering
	\includegraphics[width=\linewidth]{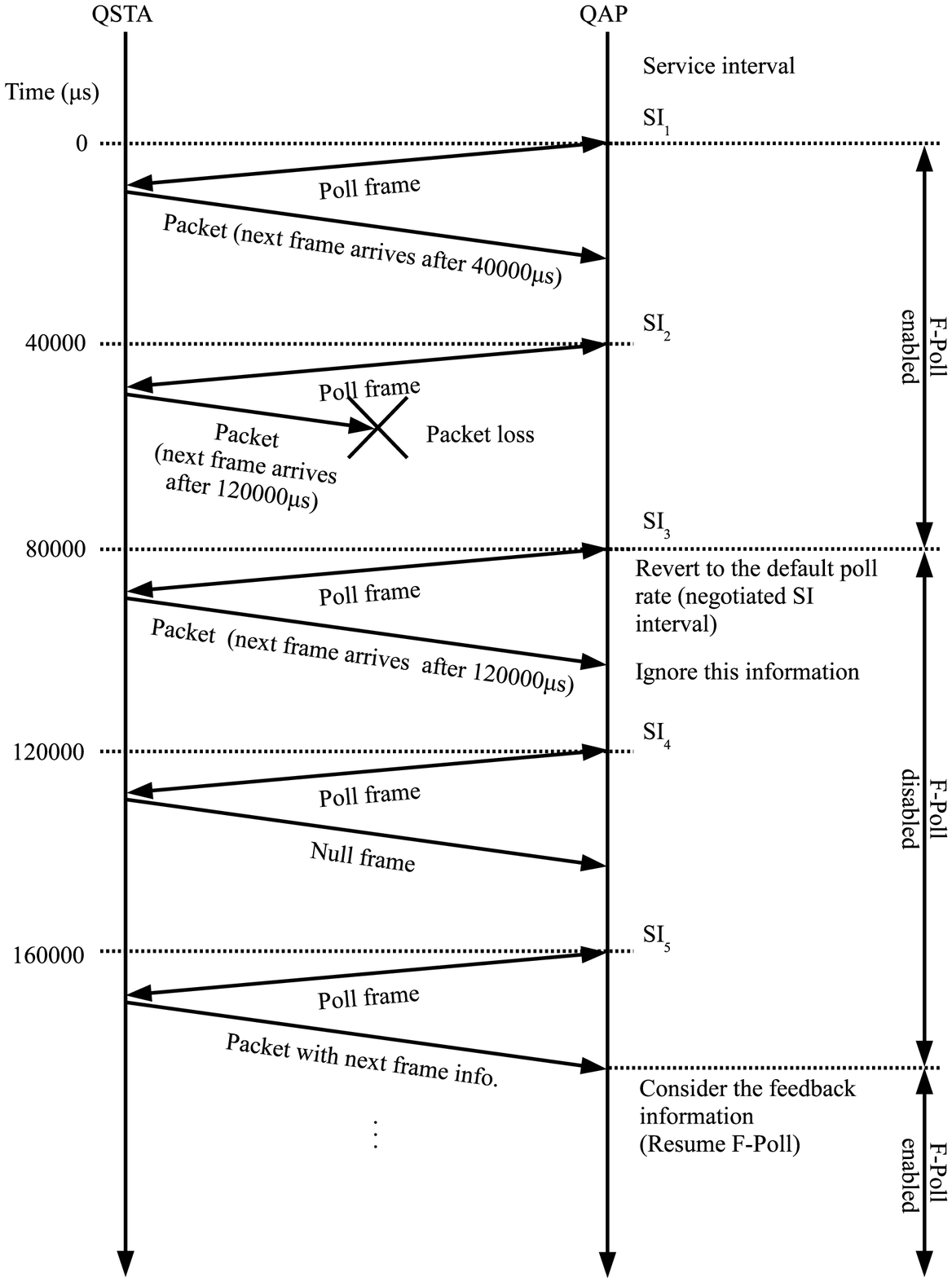}
	\caption{F-Poll Scheme Example when Packet Loss Occurs}
	\label{fig:packetlossRemedy}
\end{figure}

On the other hand, Since \gls{F-Poll} is aware about the traffic changing through maintaining the frame time of each \gls{TS} of the QSTAs, only $QSTA_{4}$, $QSTA_{5}$ will be polled in $SI_{i}$ and $QSTA_{1}$, $QSTA_{2}$ and $QSTA_{3}$ in $SI_{i+1}$. The unused channel time conserved due to this scheme will be credited to contention period of \gls{HCF}, namely \gls{EDCA} resulting remarkable reduction in the end-to-end delay of TSs of $QSTA_{4}$ and $QSTA_{5}$ in $SI_{i}$. 
Moreover, the operation of the \gls{F-Poll} when a packet loss occurs has been shown in Figure \ref{fig:packetlossRemedy}.
\section{Performance Evaluation}
\label{sec:perfEval}
In this section, \gls{F-Poll} is evaluated using simulation. The simulation setup and video traffic used for uplink traffics is described in details. Simulation has been run with different types of videos for quality measurements. The performance of the \gls{F-Poll} is compared with the \gls{HCCA} and one of the recent enhanced \gls{HCCA} scheduler, namely Enhanced \gls{EDD}. The results of the examined schemes are discussed in terms of throughput, end-to-end delay, access delay and the poll overhead.
\subsection{Simulation Setup}
\label{subsec:5_simSetup}
\begin{figure}[!h]
\centering
\includegraphics[scale=0.3]{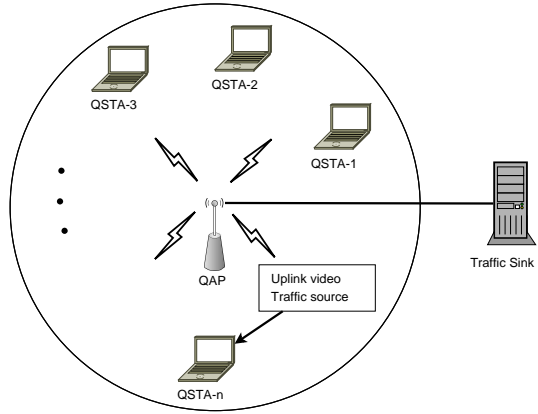}
\caption{Network Topology, One Access Point and Varying Number of Stations From One to Twelve.}
\label{fig:topology}
\end{figure}

The software implementation of the \gls{F-Poll} scheme has been developed on a network simulator with the \gls{HCCA} implementation framework \textit{ns2hcca} \cite{cicconetti2005} has been patched to provide the controlled access mode of IEEE 802.11e functions. The \textit{ns-2} Traffic Trace agent\cite{ns2Book2012} is used to generate payload bursts from the video trace file. The star topology in Figure \ref{fig:topology} has been used for constructing the simulation scenario which form an infrastructure network of one \gls{QAP} surrounded by varying number of the QSTAs ranging from 1 to 20. All QSTAs were distributed uniformly around the \gls{QAP} with a radius of 10 meters.
Stations were placed within the \gls{QAP} coverage area, in the same basic service set \gls{BSS}, and the wireless channel assumed to be ideal. Since we focus on \gls{HCCA} performance measurement, all stations operate only on the contention-free mode by setting $T_{CP}$ in Equation (\ref{eq:ACU}) to zero. \gls{QAP} is the sink receiver, while all stations are the video sources each sends only an uplink video traffic as only one flow per station is supported in \textit{ns2hcca} patch. Therefore, for simulating concurrent video streams, multiple stations are added each with one flow. 

Due to the fact that downlink TSs are maintained in \gls{QAP} queues for this reason, \gls{HC} can schedule them easily, we exclude the evaluation of the examined schemes in the presence of downlink traffics. In order to leave an ample time for initialization, stations start their transmission after 20 (sec) from the start of the simulation time and last till the simulation end. Wireless channel assumed to be an error-free. No admission control used for the sake of investigating the maximum scheduling capability of each examined algorithm under heavy traffic conditions.
Simulation parameters are summarized in Table \ref{tab:SimPars}.
\begin{table}[!h] \tabcolsep15pt
\caption {Simulation Parameters}
\centering
\begin{tabular}{p{4cm}p{3cm}}
	\hline
	Parameter 			& Value \\ \hline
	Simulation time		& 500 sec \\
	Physical layer 		& IEEE 802.11g \\
	MAC layer			& IEEE 802.11e \\
	SIFS 				& 10 $\mu s$ \\
	PIFS 				& 30 $\mu s$ \\
	Slot time 			& 20 $\mu s$ \\
	Preamble length 	& 144 bits \\
	PLCP header length 	& 48 bits \\
	PLCP Data Rate		& 1 Mbps \\
	MAC header 			& 36 bytes \\
	Physical Data rate	& 54 Mbps \\
	Basic rate 			& 6 Mbps \\ \hline
\end{tabular}
\label{tab:SimPars}
\end{table}
\subsection{Video Model Setup}
\label{subsec:5_VideoModel}
For testing the performance of the examined schemes under different traffic variability, three video sequences have been chosen from a publicly available library for video traces \cite{Fitzek2001};  Formula 1, Soccer and Mr Bean. All videos used in the simulation are H.263 format encoded at low bit rate target (16Kbps). The reason behind the selection of the low bit rate encoded videos is that although raw YUV videos are encoded at a fixed reference frame rate of 25 frames per second. The encoder skips some frames so as to achieve the target low rate. Consequently, the inter-arrival time of the frames will be higher which better represents high fluctuating \gls{VBR} traffic. Table~\ref{tab:traceStats} demonstrates some statistics of the examined traces. \gls{TSPEC} parameters used for each video traffic is shown in Table \ref{tab:VideoParas} with regards to video \gls{QoS} requirements.
\begin{table}[!h]
\centering
\caption {Frame Statistics of MPEG--4 Video Trace Files \cite{Fitzek2001}}
\begin{tabular}{ p{5cm}lll }
	\hline
	Parameter				& Formula 1	& Soccer	& Mr Bean	\\ \hline
	Comp. ratio (YUV:H263)	& 476.07	& 476.30	& 476.38	\\
	Mean size (byte)		& 519		& 655		& 403		\\
	Maximum size (byte)		& 4831		& 4647		& 3265		\\
	CoV of bit rate			& 0.21		& 0.17		& 0.34		\\
	Mean bit rate (bit/sec)	& 1.6e+04	& 1.6e+04	& 1.6e+04	\\
	Peak bit rate (bit/sec)	& 7.8e+04	& 7.3e+04	& 9.7e+04	\\ \hline
\end{tabular}
\label{tab:traceStats}
\end{table}
\begin{table}[!h]
\centering
\caption {Traffic Parameters for Video Streams.}
\begin{tabular}{llll}
	\hline
	Parameter				& Formula 1		& Soccer	& Mr Bean		\\ \hline
	Nominal MSDU Size		& 519 bytes		& 655  bytes& 403 bytes		\\
	Maximum MSDU Size		& 4831 bytes	& 4647 bytes& 3265 bytes	\\
	Mean Data Rate			& 16 Kbps		& 16 Kbps	& 16 Kbps		\\
	Peak Data Rate			& 78 Kbps		& 73 Kbps	& 97 Kbps		\\
	Delay Bound				& 0.08 sec		& 0.08 sec	& 0.08 sec		\\
	Minimum Physical Rate	& 54 Mbps		& 54 Mbps	& 54 Mbps		\\
	Maximum Service Interval& 0.04 sec		& 0.04 sec	& 0.04 sec		\\ \hline
\end{tabular}
\label{tab:VideoParas}
\end{table}
\subsection{Results and Discussions}
\label{subsec:5_results}
In our simulation study the aim was to improve the \gls{QoS} provision of the \gls{HCCA} algorithm. Simulation has been run to demonstrate the performance of the examined schemes with the same simulation scenario. The main objective is to achieve better \gls{QoS} support by avoiding polling stations that have no data backlogged at their transmission queues. Packet end-to-end delay of the uplink traffics has been evaluated in this research which considered as one of the important metrics for measuring \gls{QoS} efficiency for supporting multimedia applications such as video streams. In order validate the behavior of the examined schemes, the measurements have been done with increasing number of TSs. System throughput was also investigated to verify that the improvement in delay is achieved without jeopardizing the wireless channel efficiency.
\subsubsection{Mean Access Delay Analysis}
\begin{figure}[t!]
	\centering
	\subfigure[Formula1 Video]{
		\label{fig:accessDly1}
		\includegraphics[angle=-90,scale=0.22]{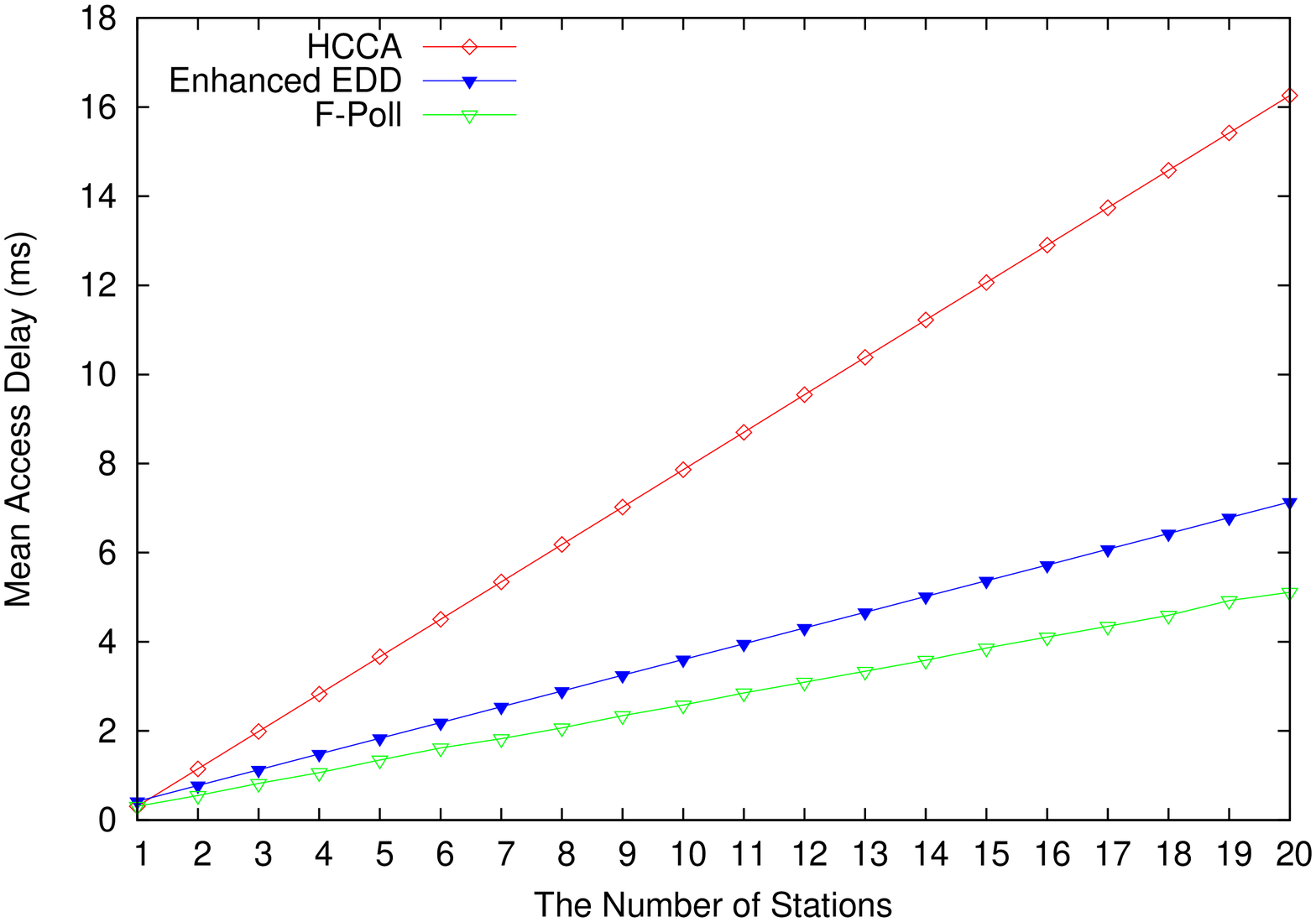}
	}
	\subfigure[Soccer Video]{
		\label{fig:accessDly2}
		\includegraphics[angle=-90,scale=0.22]{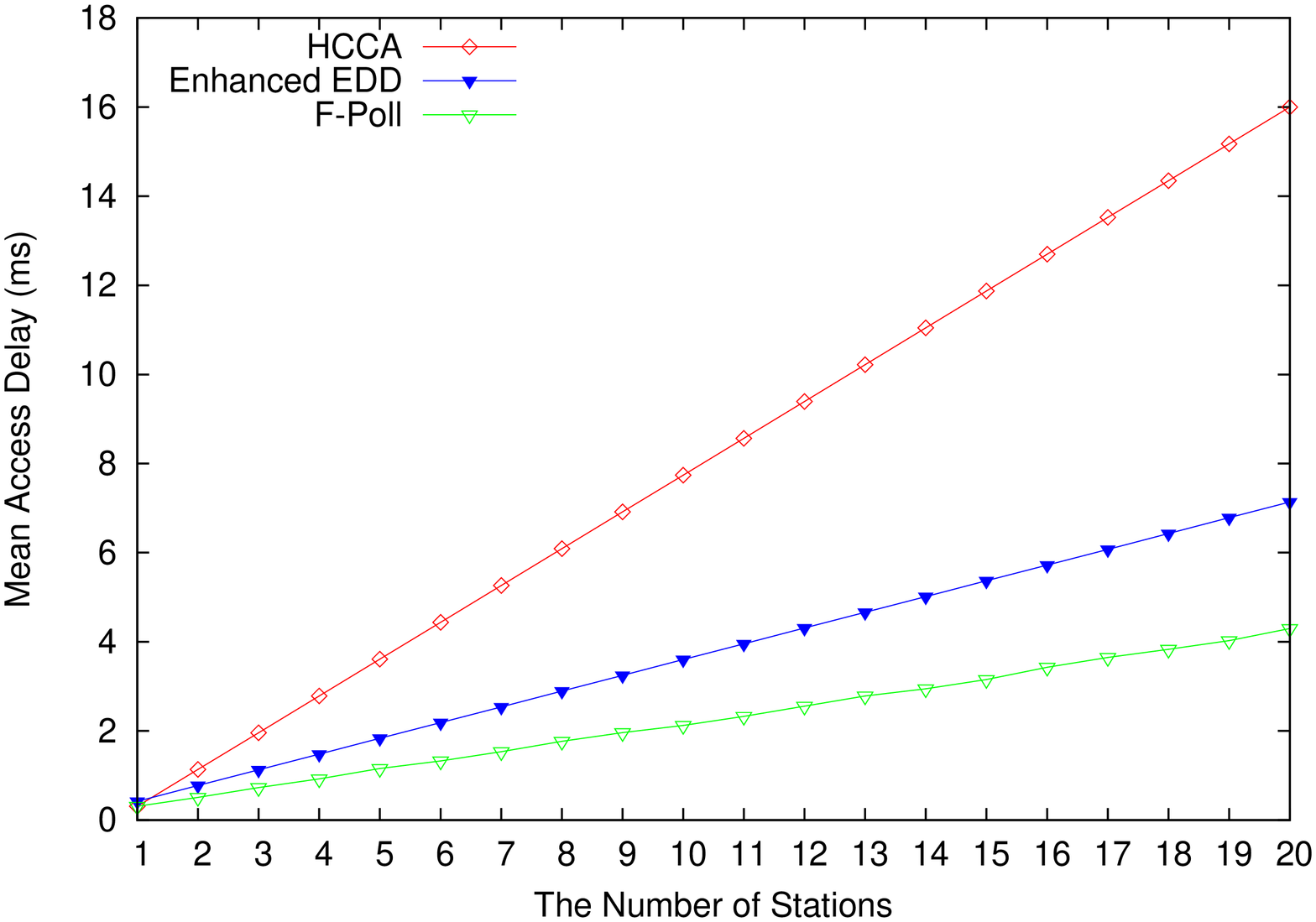}
	}
	\subfigure[Mr Bean Video]{
		\label{fig:accessDly3}
		\includegraphics[angle=-90,scale=0.22]{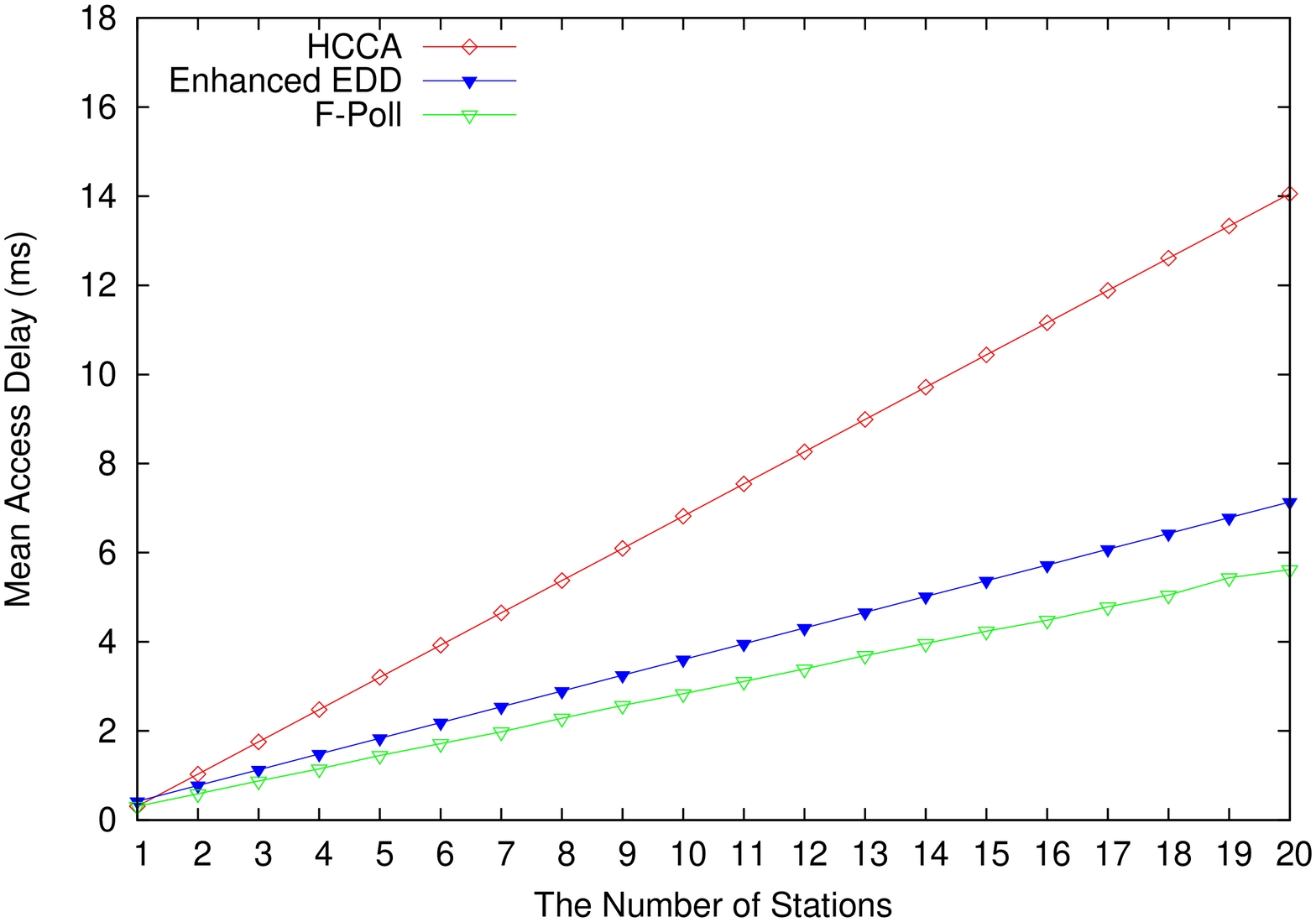}
	}
	\caption{Packet Mean Access Delay as a Function of The Number of Stations.}
	\label{fig:accessDly}
\end{figure}

The packet access delay is referred to the time taken from the packet generation at the station until it's been transmitted from the \gls{MAC} layer. The mean access delay is calculated in Equation (\ref{eq:meanDelay5})
\begin{eqnarray}
\label{eq:meanDelay5}
MeanAccessDelay  = \frac {1}{N} \sum_{i=1}^{N} ( S_{i}-G_{i}),
\end{eqnarray}
where $G_{i}$ is the generation time of packet $i$ from the source station, $S_{i}$ is the sending time of the particular packet (i) from \gls{MAC} layer of the station and $N$ is the total number of packets for all flows in the system. This metric reflects the delay time from a station being ready to transmit until it being served. In this experiment, we illustrate the efficiency of the \gls{F-Poll} over \gls{HCCA} polling and Enhanced \gls{EDD} schemes in adapting to the fast changing of \gls{VBR} traffic over the time. 
Figures \ref{fig:accessDly1}, (b) and (c)  show the average access delay of the data packets for the examined video traces. The proposed \gls{F-Poll} scheme exhibits low access delay in all videos compared to both \gls{HCCA} polling scheme and Enhanced \gls{EDD}. The reason behind this low delay is that the \gls{F-Poll} scheme is always aware about the change in uplink traffic profile and only polls the stations in need. Thus, the polls overhead is minimized which in turn the shorten the time the packets wait in transmission queues. Note that Enhanced \gls{EDD} accelerates the mSI according to the transmission queue status and the average \gls{TXOP} assignment. This concept can minimize the delay when traffic shows variability in packet size only. However, in the case of traffics that show variability in packet inter-arrival time the delay caused by the wasted polls still persists since it is not being addressed. In contrary, \gls{HCCA} polling scheme polls all stations regardless their actual needs which causes an increase in the queuing time awaiting a poll message. The maximum mean access delay experienced using the \gls{F-Poll} scheme was as low as 5, 9 and 6 ms whereas in the reference poll scheme was 19, 14 and 20 ms for Formula 1, Soccer and Mr Bean video sequences respectively. 

One can see the acute increase of the access delay by increasing the network load in the case of the \gls{HCCA} polling scheme which can be justified by the high increase of the poll overhead in the network. The Enhanced \gls{EDD} achieved up to 58\% over \gls{HCCA} in the three videos whereas \gls{F-Poll} has achieved around 27\% over Enhanced \gls{EDD} for Formula 1 and Mr Bean videos whereas, for Soccer video, \gls{F-Poll} has enhanced the delay by up to 38\% over Enhanced \gls{EDD}.
\subsubsection{End-to-end Delay Analysis}
\label{subsubsec:5_e2eD}
\begin{figure*}
	\centering
	\subfigure[HCCA]{
		\label{fig:e2eDly1ref}
		\includegraphics[scale=0.4,angle=-90]{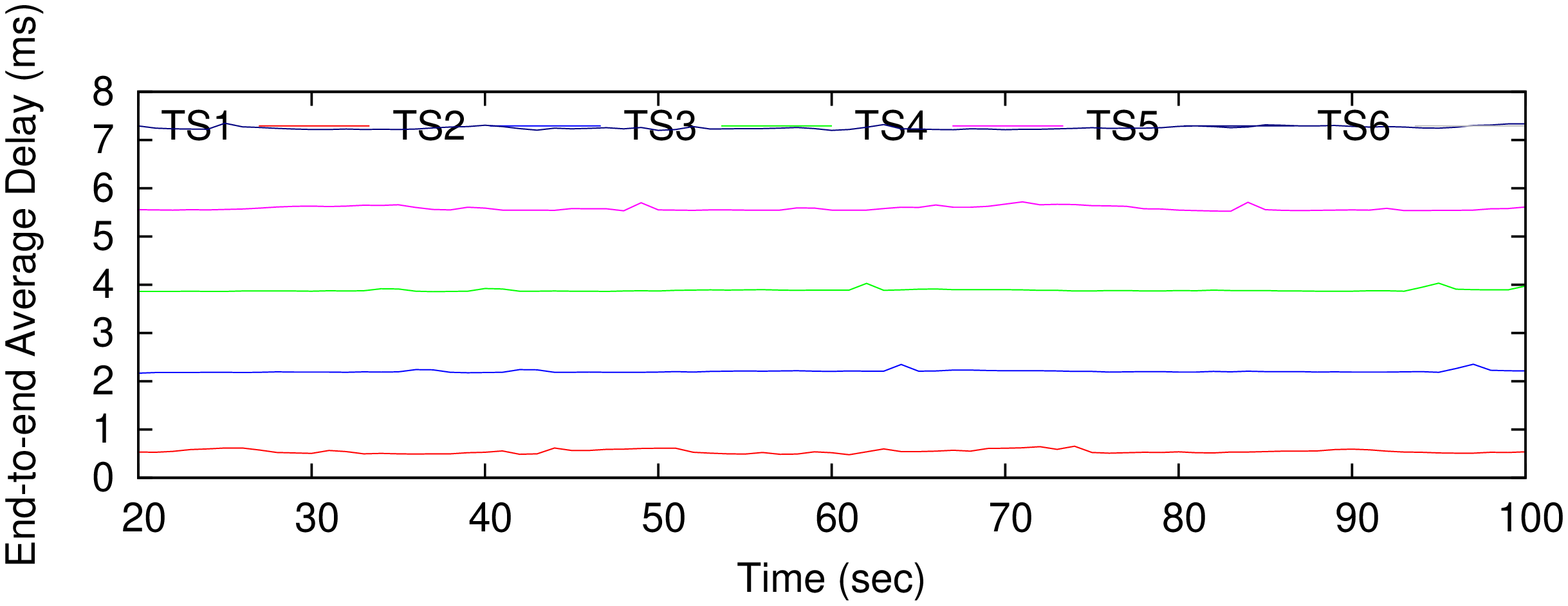}
	}
	\subfigure[Enhanced EDD]{
		\label{fig:e2eDly1edd}
		\includegraphics[scale=0.4,angle=-90]{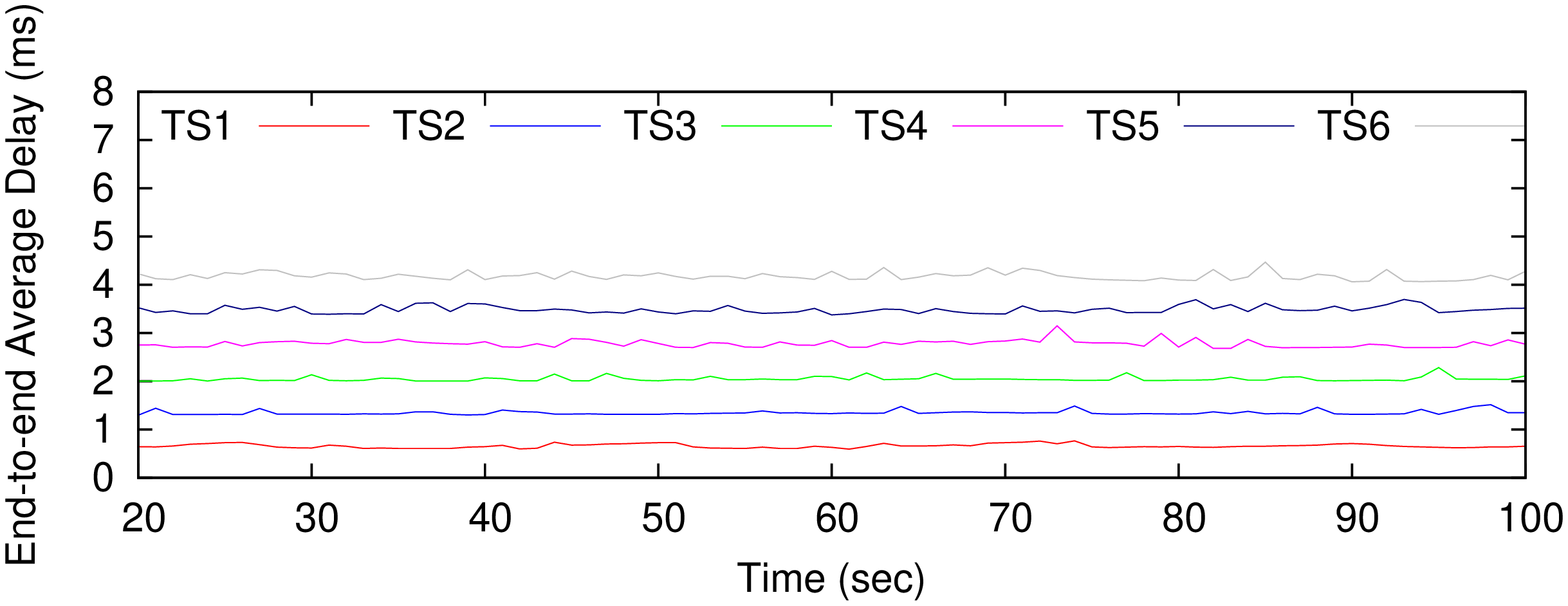}
	}
	\subfigure[F-Poll]{
		\label{fig:e2eDly1Fpoll}
		\includegraphics[scale=0.4,angle=-90]{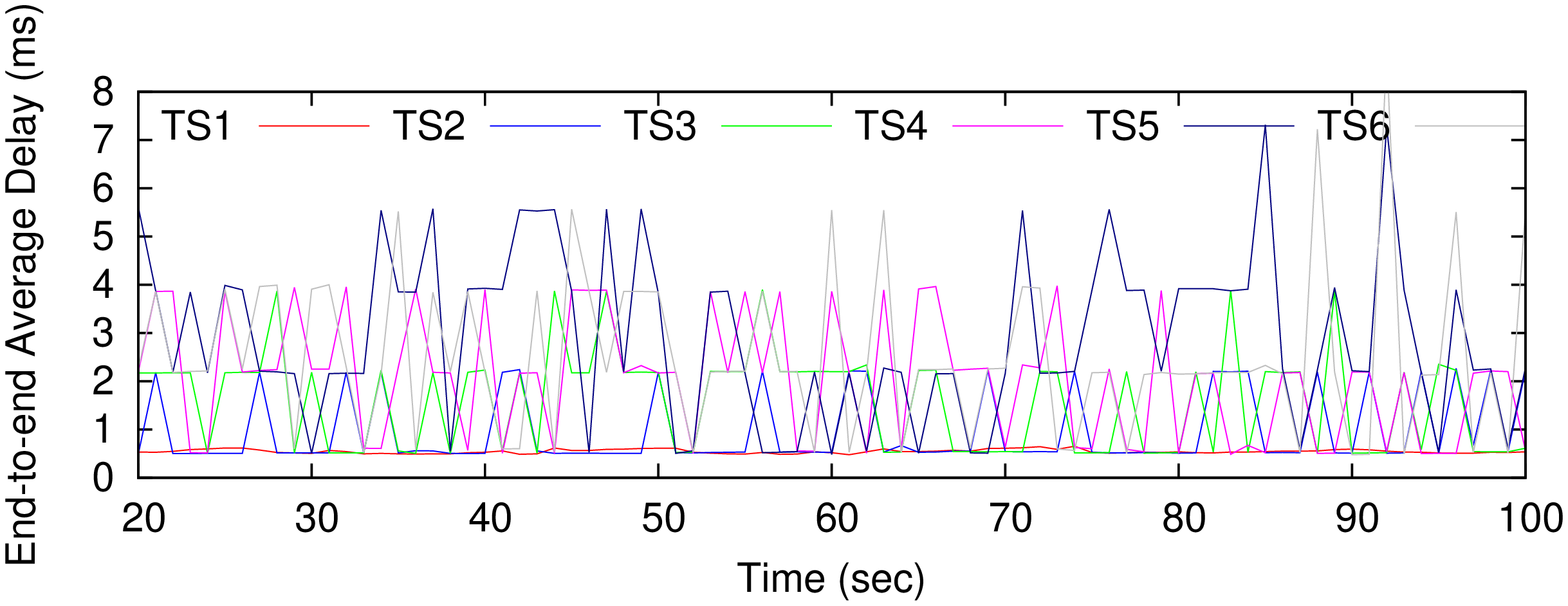}
	}
	\caption{Packet End-to-end Delay for Formula1 Video}
	\label{fig:Pe2eDlyFormula1}
\end{figure*}

\begin{figure}
	\centering
	\subfigure[HCCA]{
		\label{fig:e2eDly2ref}
		\includegraphics[scale=0.4,angle=-90]{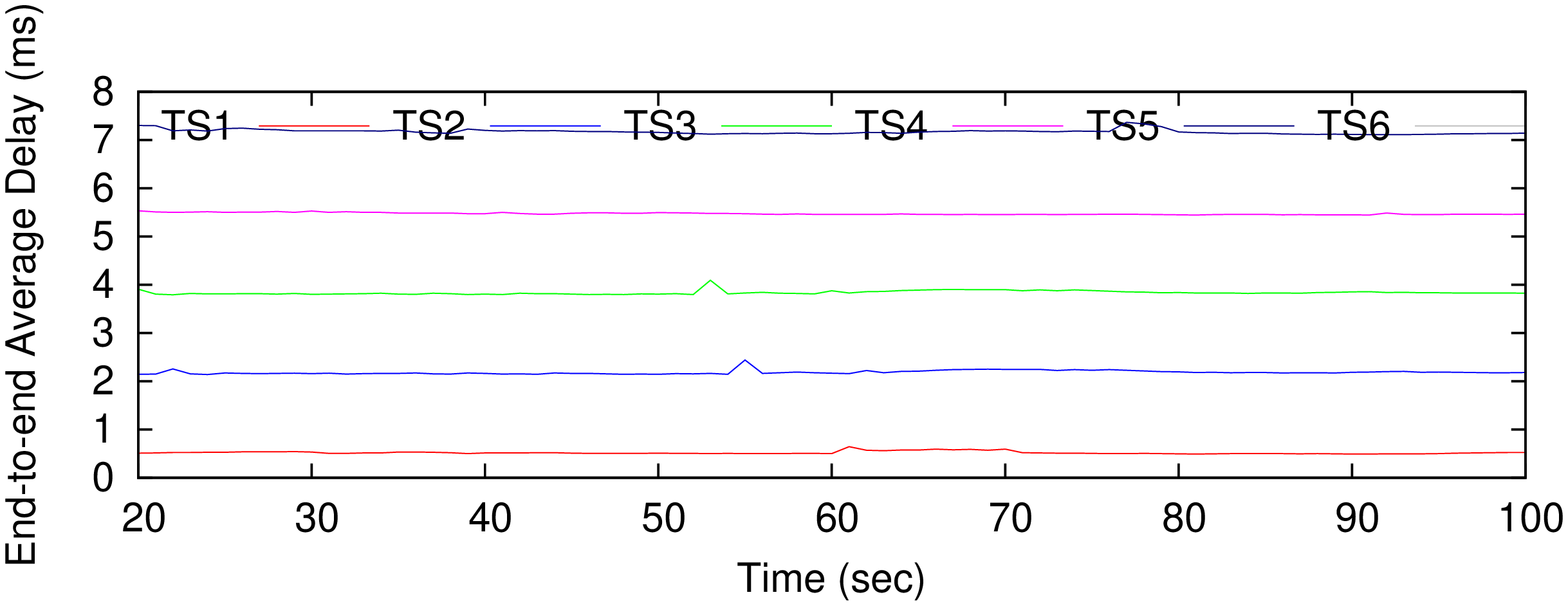}
	}
	\subfigure[Enhanced EDD]{
		\label{fig:e2eDly2edd}
		\includegraphics[scale=0.4,angle=-90]{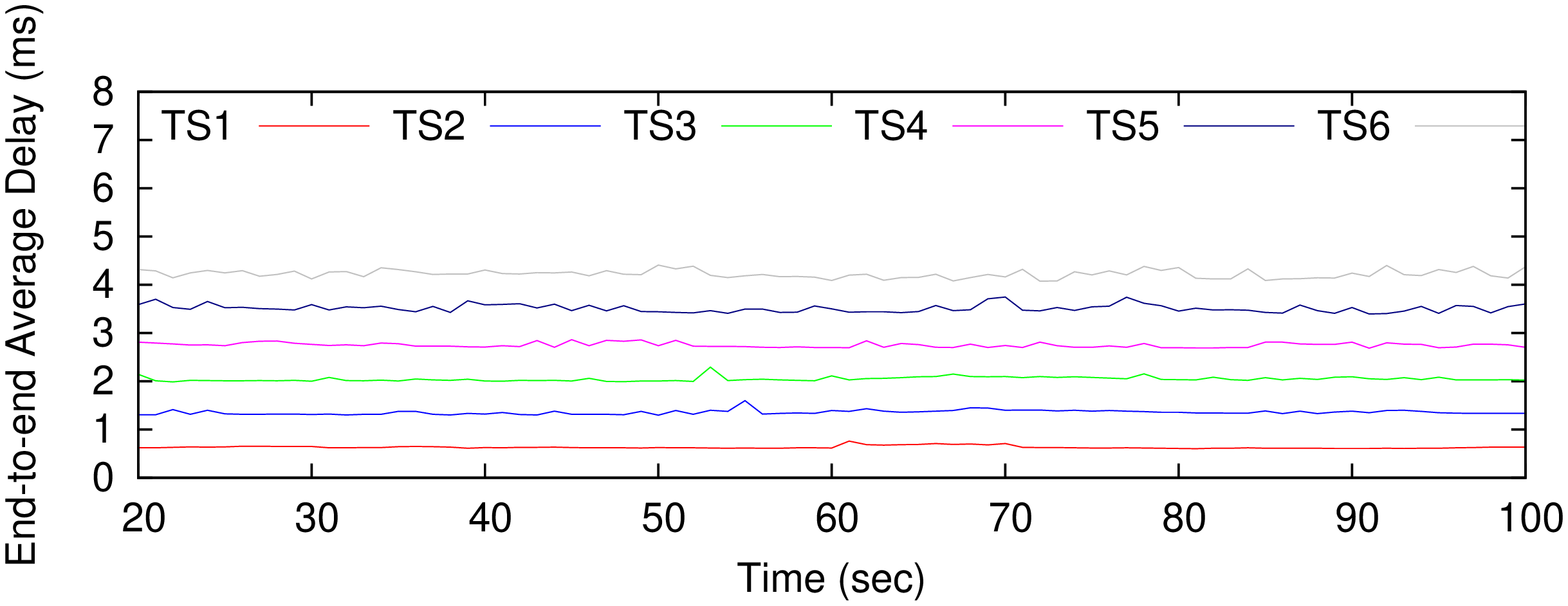}
	}
	\subfigure[F-Poll]{
		\label{fig:e2eDly2Fpoll}
		\includegraphics[scale=0.4,angle=-90]{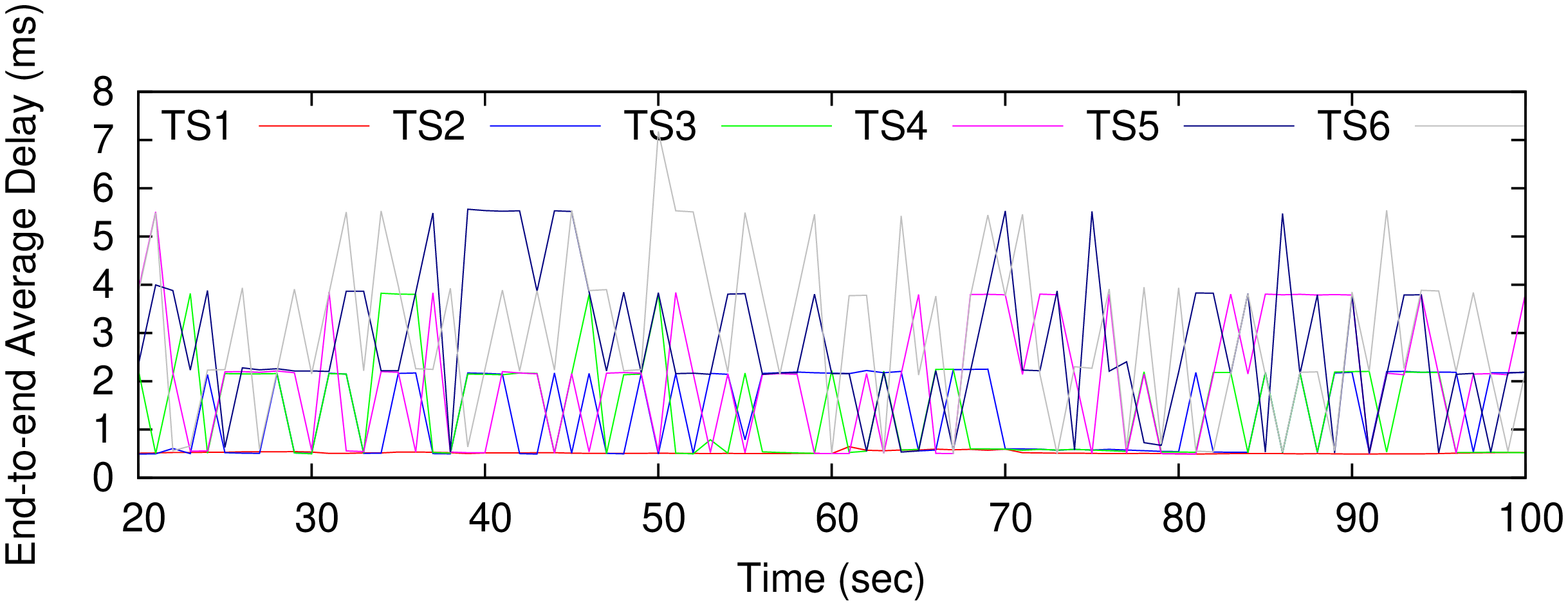}
	}
	\caption{Packet End-to-end Delay for Soccer Video}
	\label{fig:Pe2eDlySoccer}
\end{figure}

\begin{figure}[!h]
	\centering
	\subfigure[HCCA]{
		\label{fig:e2eDly3ref}
		\includegraphics[scale=0.4,angle=-90]{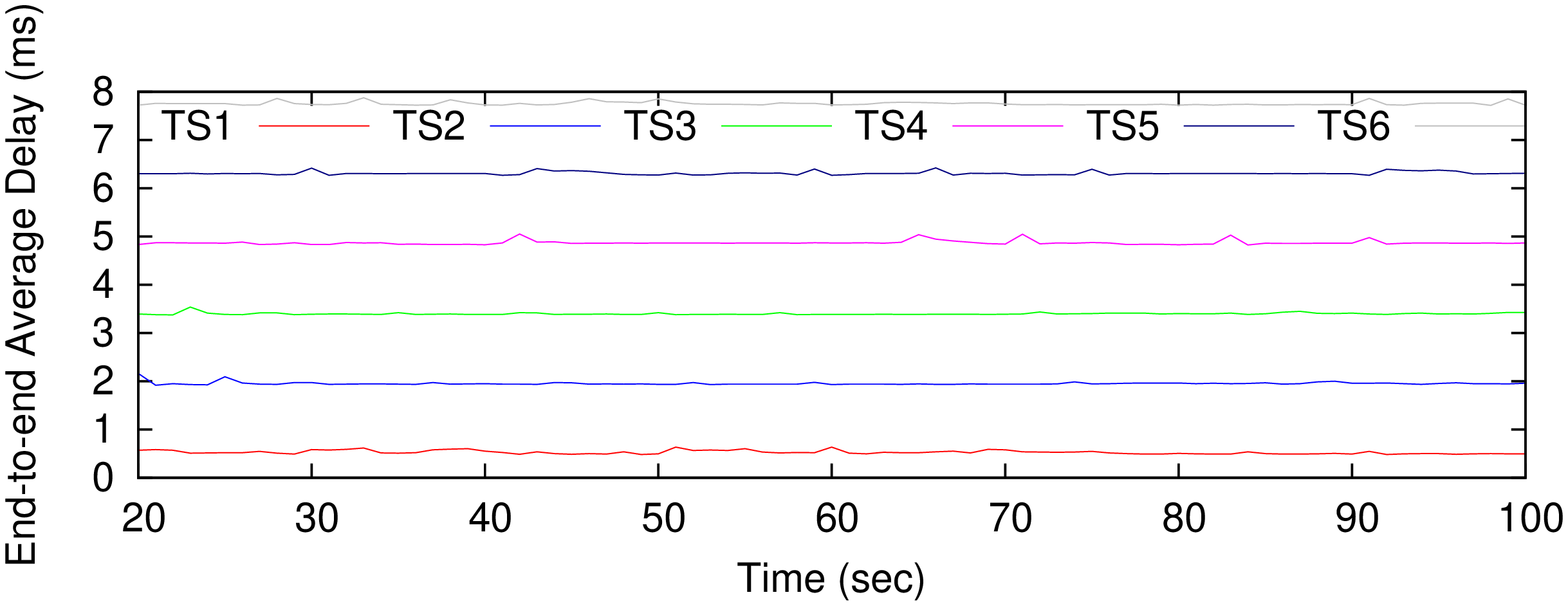}
	}
	\subfigure[Enhanced EDD]{
		\label{fig:e2eDly3edd}
		\includegraphics[scale=0.4,angle=-90]{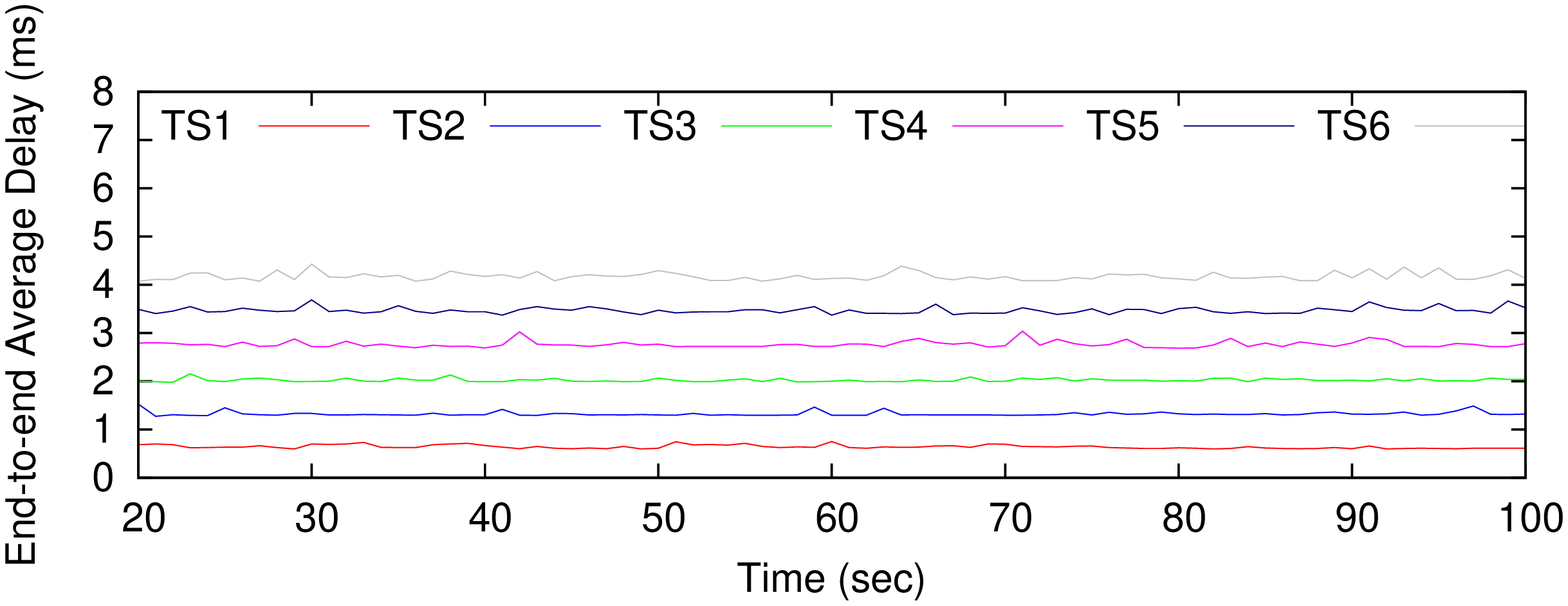}
	}
	\subfigure[F-Poll]{
		\label{fig:e2eDly3Fpoll}
		\includegraphics[scale=0.4,angle=-90]{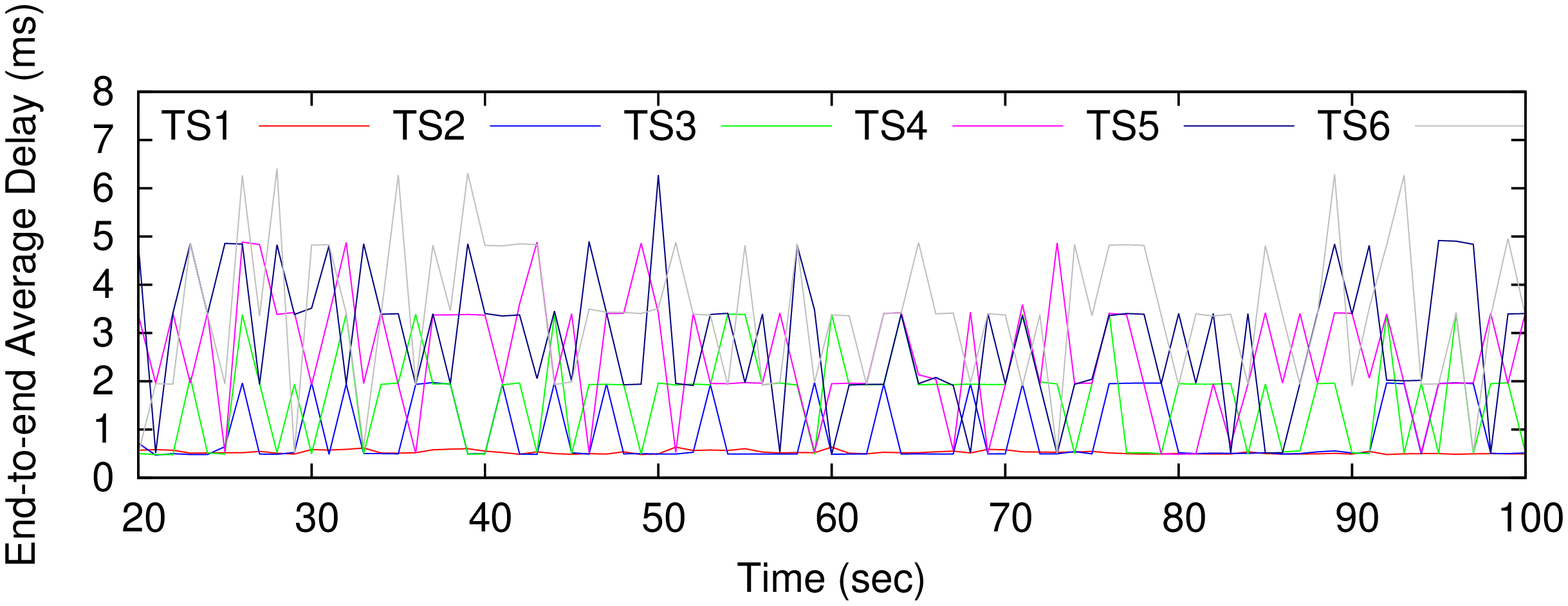}
	}
	\caption{Packet End-to-end Delay for Mr Bean Video}
	\label{fig:Pe2eDlyMrBean}
\end{figure}

The end-to-end delay is defined as the time elapsed from the generation of the packet in the application layer of the source \gls{QSTA} until it's been received at the sink node, \gls{QAP}. The end-to-end delay of the examined schemes has been measured using the three videos to study their efficiency over different traffic variability. Figures~\ref{fig:Pe2eDlyFormula1}, \ref{fig:Pe2eDlySoccer} and \ref{fig:Pe2eDlyMrBean} illustrate the effect of poll overhead on the delay of the packets of different TSs of 6 QSTAs in the system and for a duration of 100 seconds. Figures \ref{fig:e2eDly1ref}, \ref{fig:e2eDly2ref} and \ref{fig:e2eDly3ref} depict the delay experienced by data packets when using the legacy round-robin scheme of \gls{HCCA} for each traffic stream. Figures \ref{fig:e2eDly1edd}, \ref{fig:e2eDly2edd} and \ref{fig:e2eDly3edd} reveals that Enhanced \gls{EDD} has reduced the delay for all TSs yet it behave similar to \gls{HCCA} as the delay of each \gls{TS} remain the same during traffic lifetime as it does not address the issue of the over polling. Since traffics are \gls{VBR}, in some SIs, TSs packets are likely to wait for all unready stations prior in the polling list to the particular station to reply null frames. Consequently, each TSs are likely to have near the same delay each \gls{SI}. This issue becomes worse when the polling list increases where the stations in the last exposed to higher delay. It is worth noting  that the preceding TSs are most likely to experience lower delay, in this case TS1 has the lowest packet end-to-end delay while TS6 shows highest delay among all video streams.

Figures \ref{fig:e2eDly1Fpoll}, \ref{fig:e2eDly2Fpoll} and \ref{fig:e2eDly3Fpoll} exhibit that \gls{F-Poll} scheme succeeded in minimizing the delay for all TSs and the reason is that our scheme minimized the poll overhead by avoiding poll unready stations. The fluctuation in the delay of each \gls{TS} is subject to the number of transmitting stations ahead to that station in each \gls{SI}. In general, the results demonstrate that the \gls{F-Poll} scheme minimize the end-to-end delay even in the case of high network load.

\subsubsection{Poll Overhead Ratio}
\label{subsubsec:5_pollOvhd}
\begin{figure}[!h]
	\centering
	\subfigure[Formula1 Video]{
		\label{fig:polloverhead1}
		\includegraphics[angle=-90,scale=0.22]{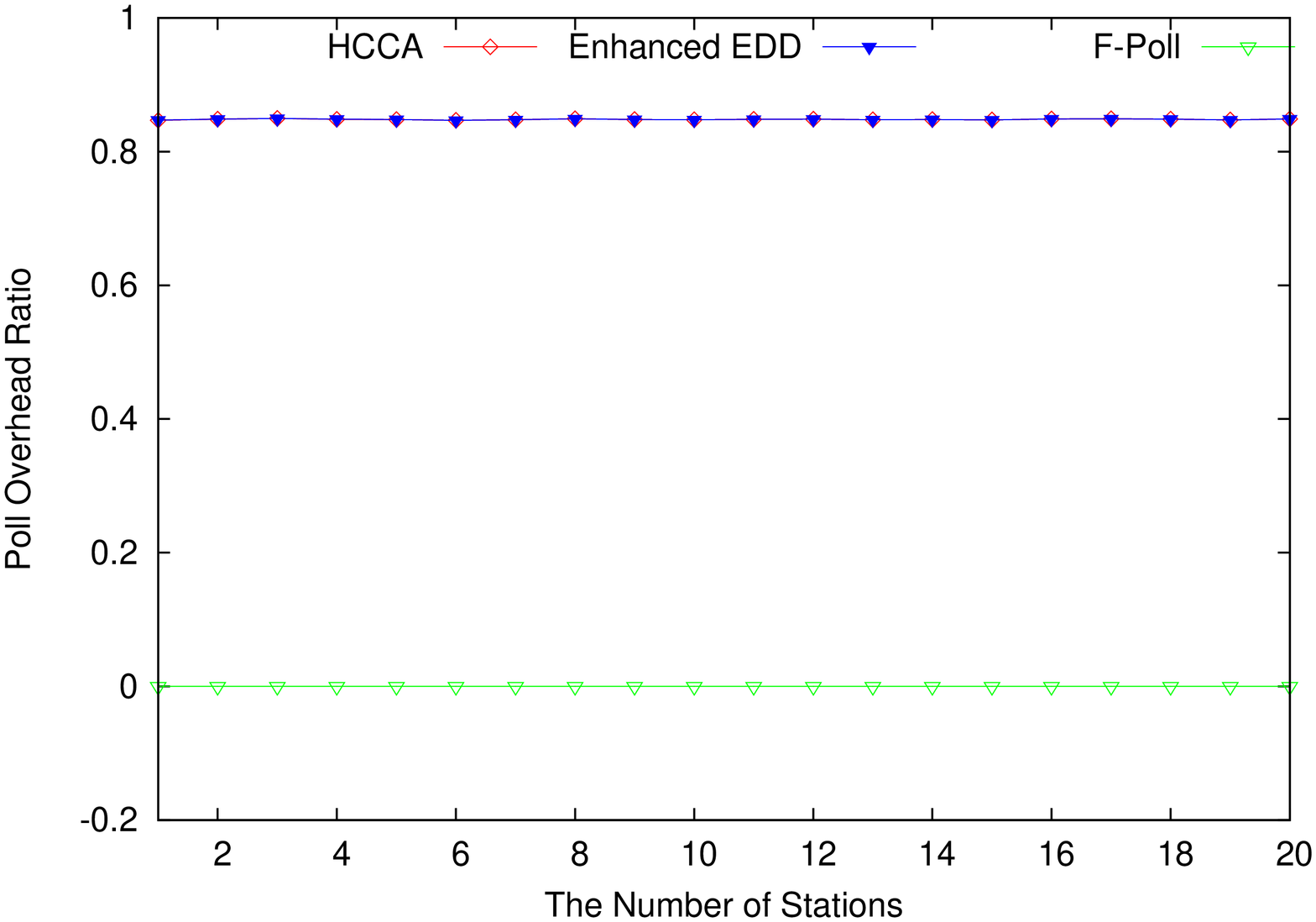}
	}
	\subfigure[Soccer Video]{
		\label{fig:polloverhead2}
		\includegraphics[angle=-90,scale=0.22]{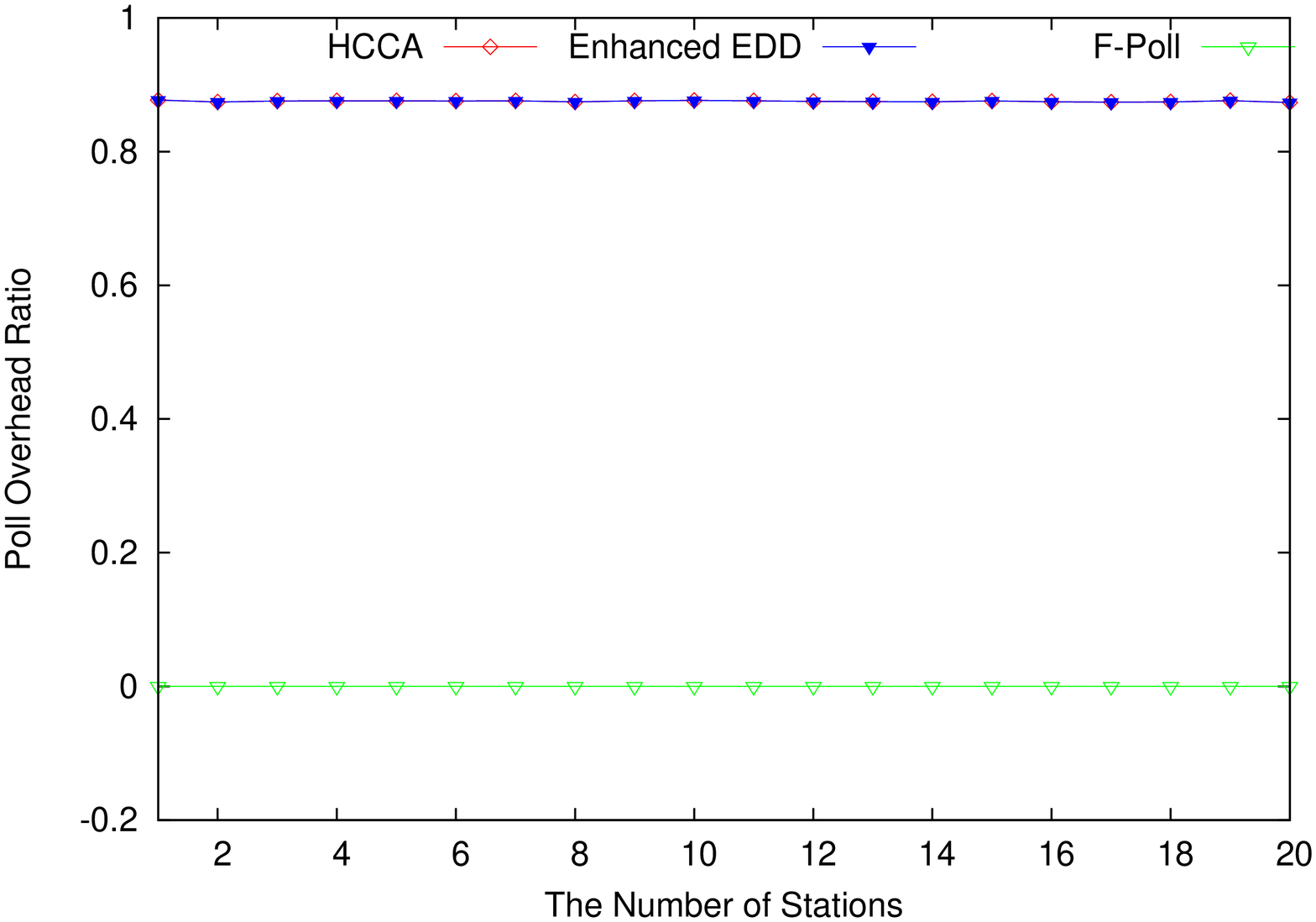}
	}
	\subfigure[Mr Bean Video]{
		\label{fig:polloverhead3}
		\includegraphics[angle=-90,scale=0.22]{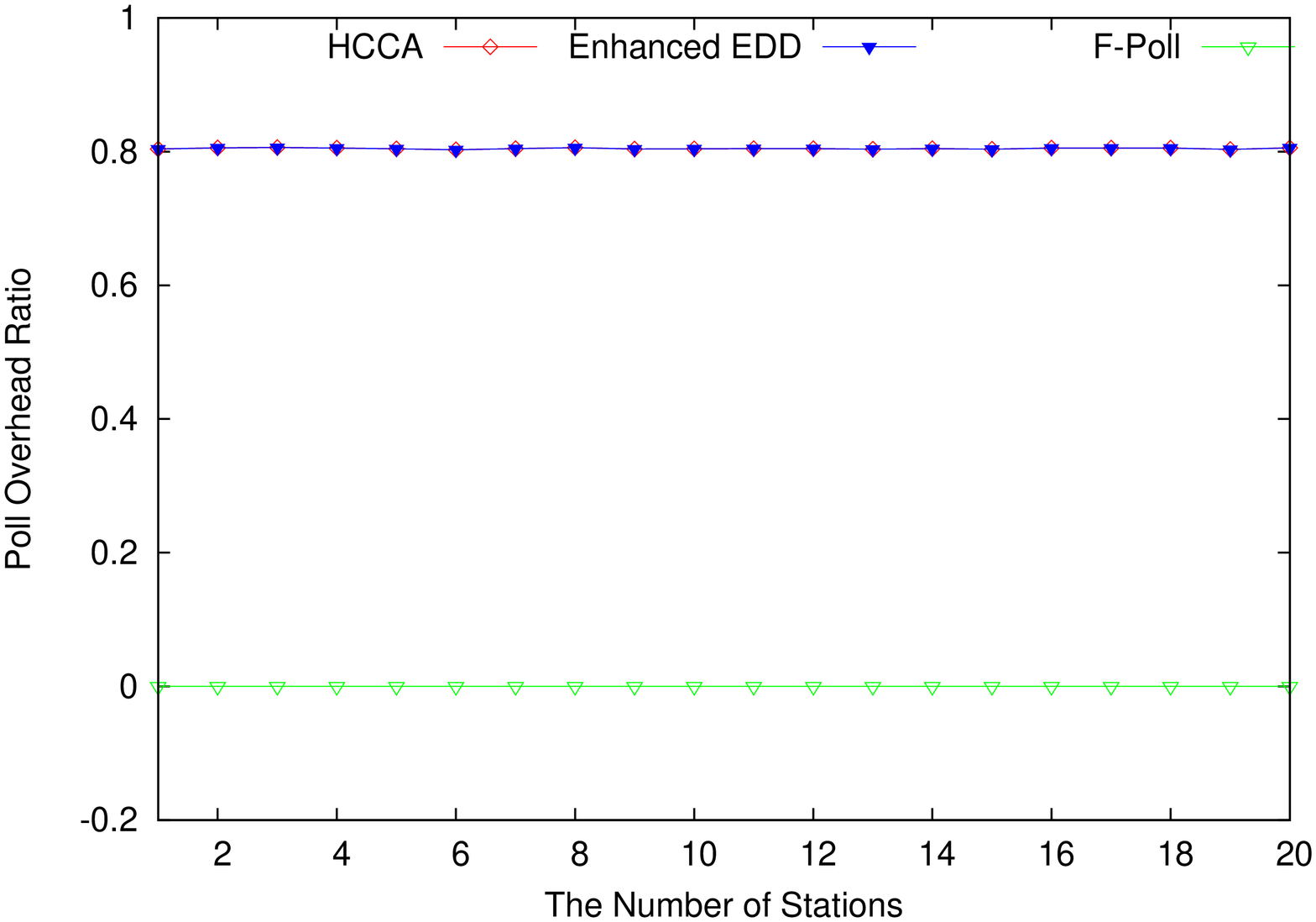}
	}
	\caption{Poll Overhead as a Function of The Number of Stations.}
	\label{fig:polloverhead}
\end{figure}

Figures \ref{fig:polloverhead1},  (b) and (c) illustrate the overpolling issue illustrated in section \ref{sec:preStudy} for the studied video traces. Poll overhead ratio is calculated as the ratio of the number of Null-frames to the number of poll frame sent to the uplink TSs. Since the studied videos show high variability in their profile which lead to aggravate the Null-frame responses. In the reference \gls{HCCA} scheme, the poll overhead ratio reaches about 85\% for all the studied video traces with despite that the network is heavily loaded or not. F-Poll scheme as low as zero since the scheduler aware about the exact time for the next arrival packet.
\\
\subsubsection{Number of Polls Versus Number of Packets}
\begin{figure}[!h]
	\centering
	\subfigure[Formula1 Video]{
		\label{fig:pollsvspkt1Ref}
		\includegraphics[angle=-90,scale=0.22]{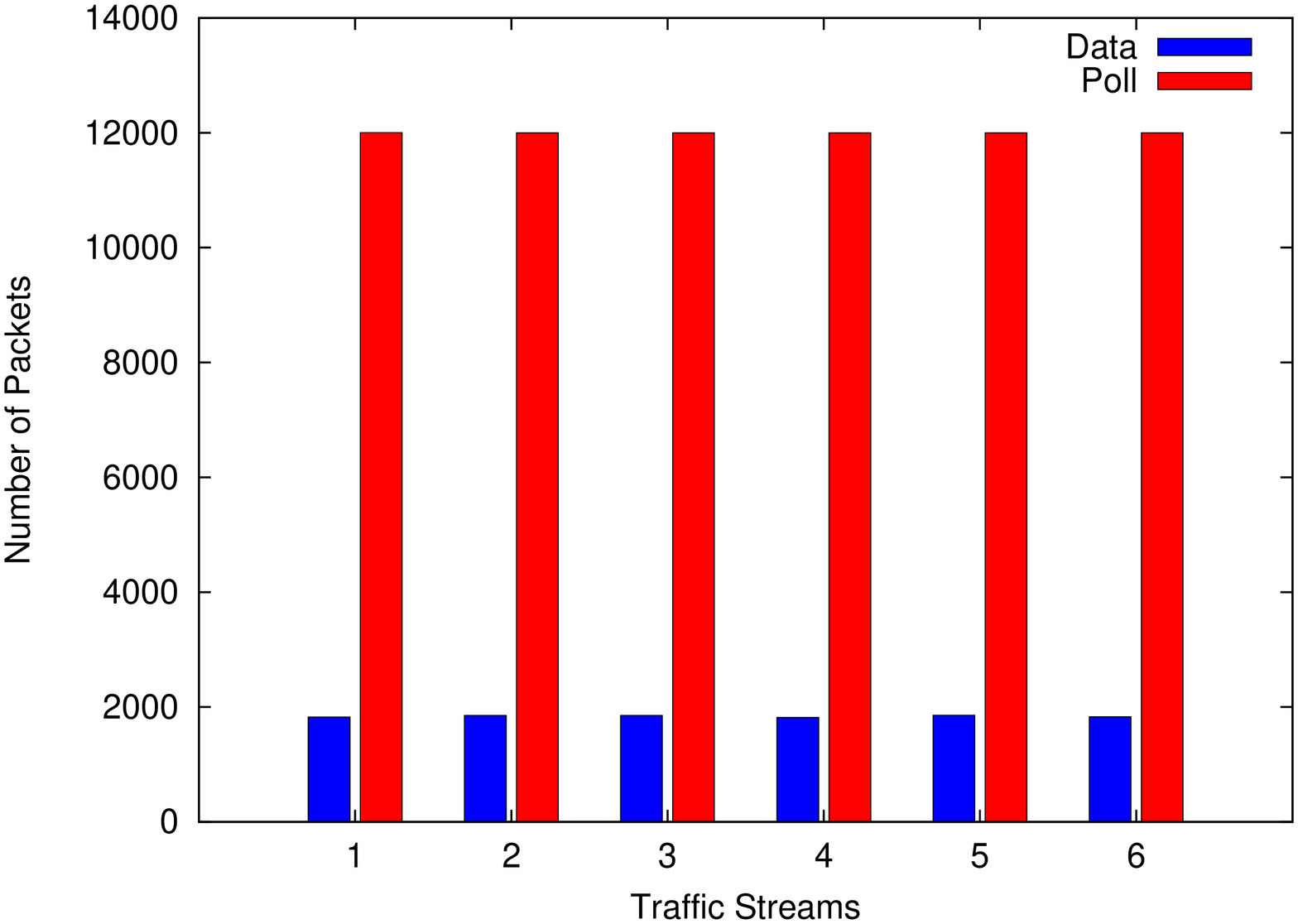}
	}
	\subfigure[Soccer Video]{
		\label{fig:pollsvspkt2Ref}
		\includegraphics[angle=-90,scale=0.22]{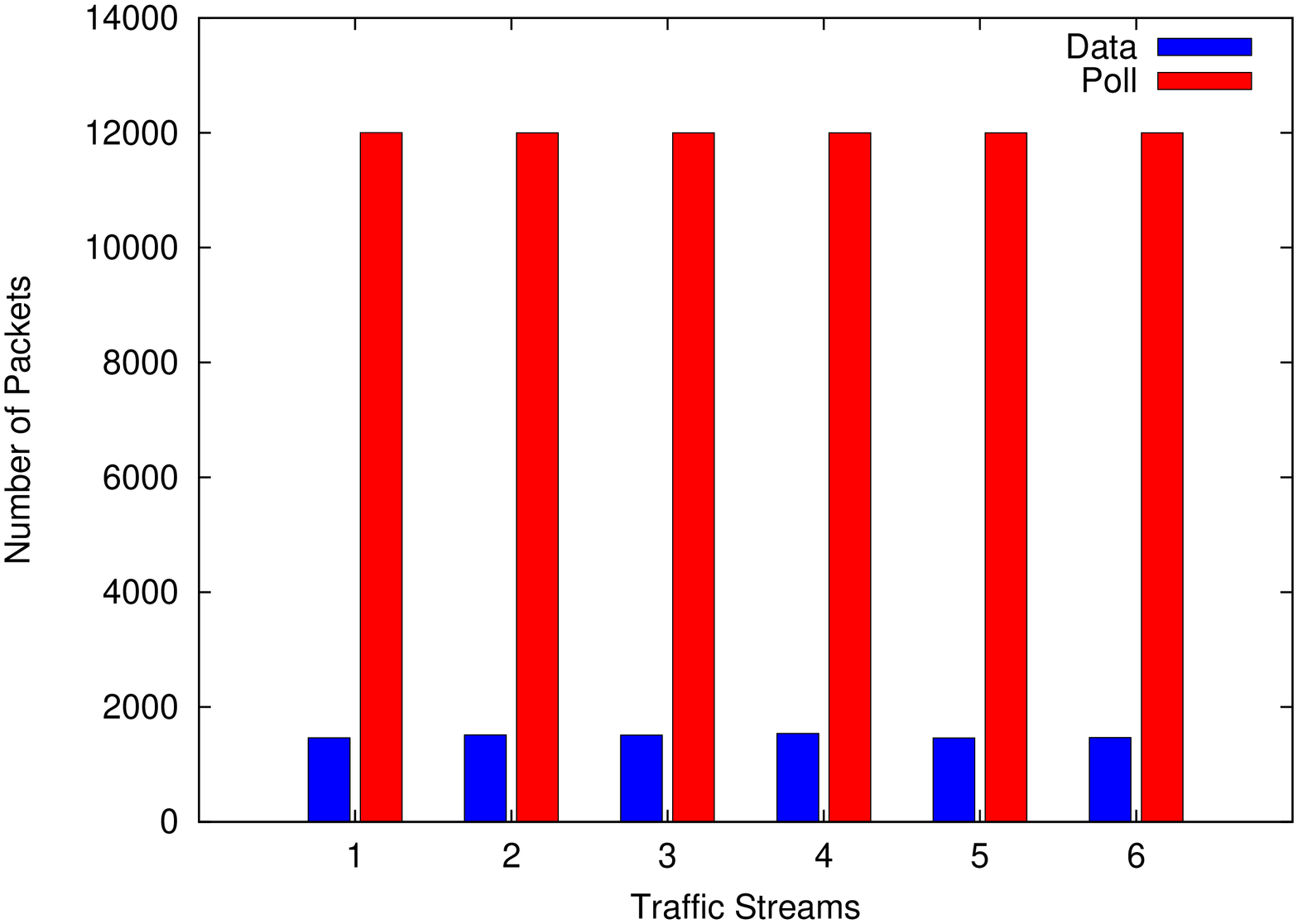}
	}
	\subfigure[Mr Bean Video ]{
		\label{fig:pollsvspkt3Ref}
		\includegraphics[angle=-90,scale=0.22]{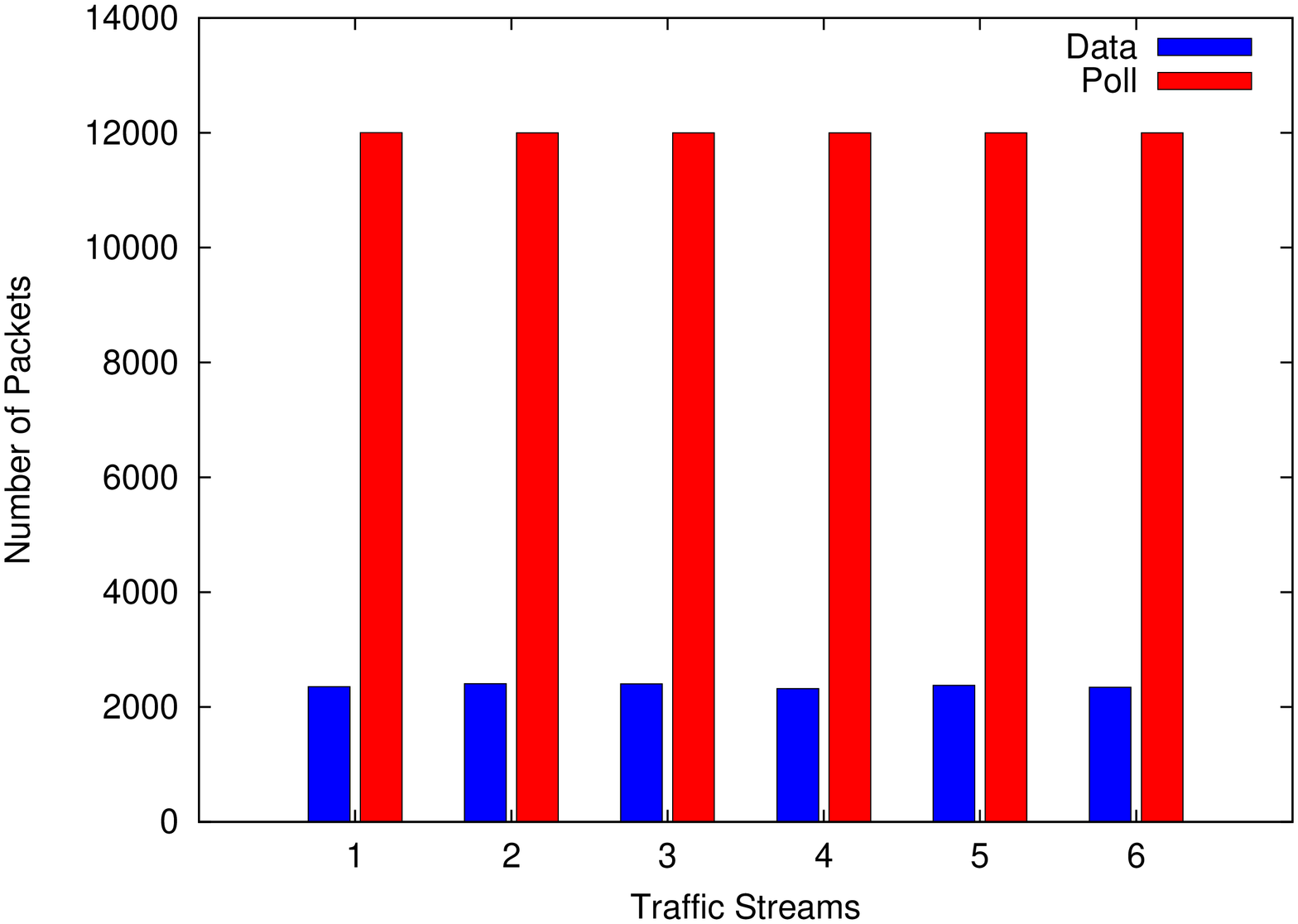}
	}
	\caption{Polls vs Packets of HCCA as a Function of The Number of Stations.}
	\label{fig:pollsvspktHCCA}
\end{figure}

\begin{figure}[!h]
	\centering
	\subfigure[Formula1 Video]{
		\label{fig:pollsvspkt1edd}
		\includegraphics[angle=-90,scale=0.22]{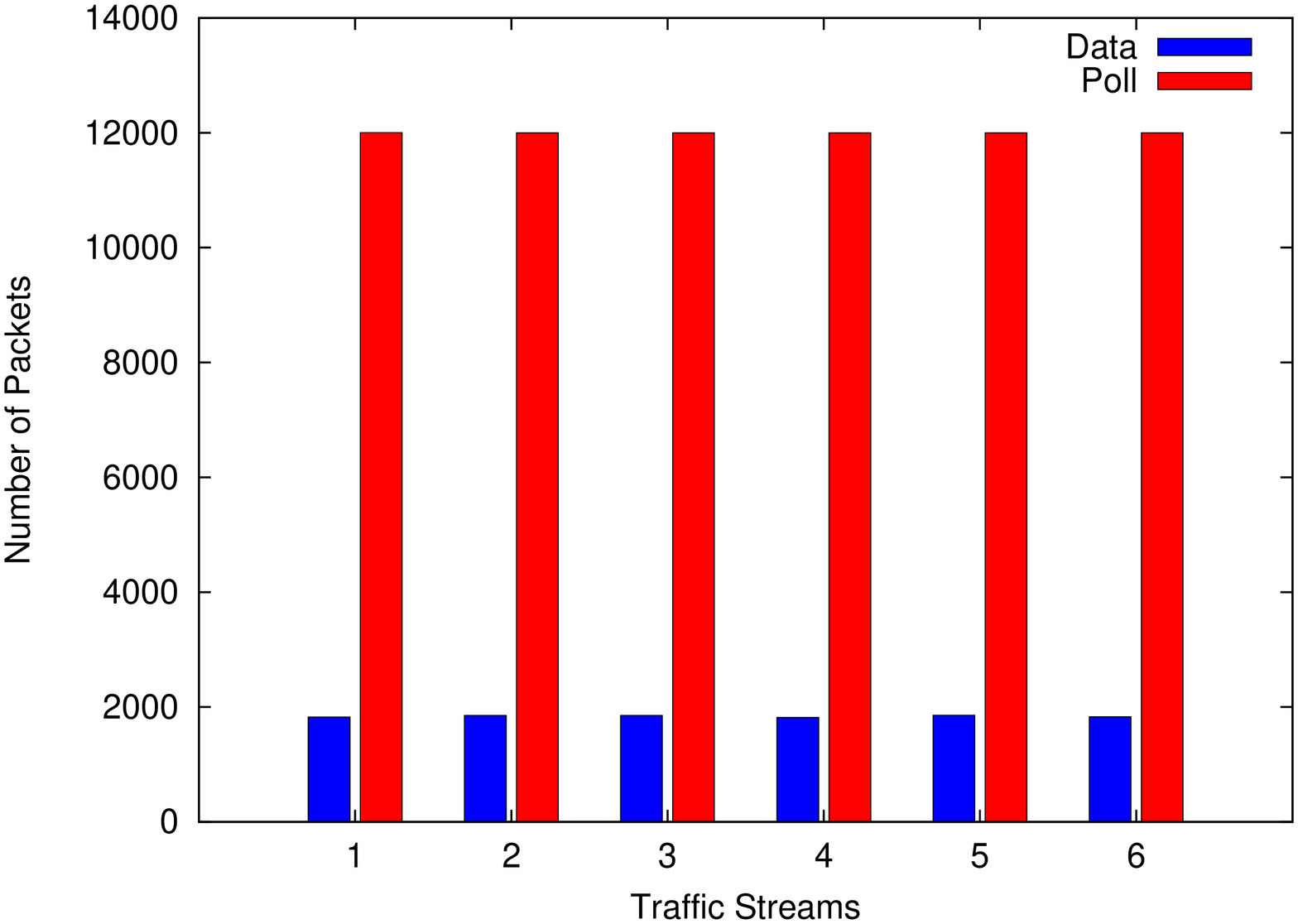}
	}
	\subfigure[Soccer Video]{
		\label{fig:pollsvspkt2edd}
		\includegraphics[angle=-90,scale=0.22]{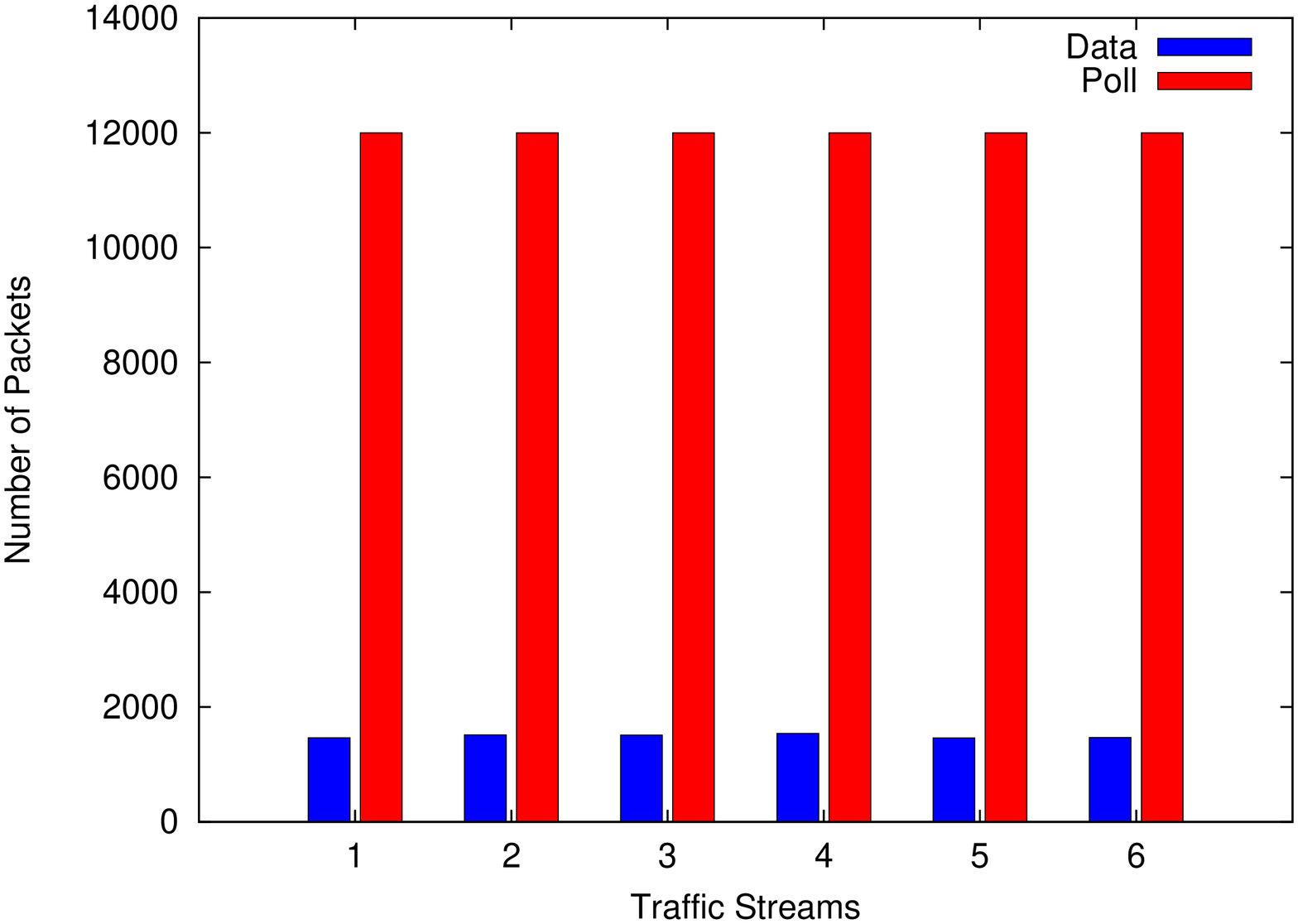}
	}
	\subfigure[Mr Bean Video]{
		\label{fig:pollsvspkt3edd}
		\includegraphics[angle=-90,scale=0.22]{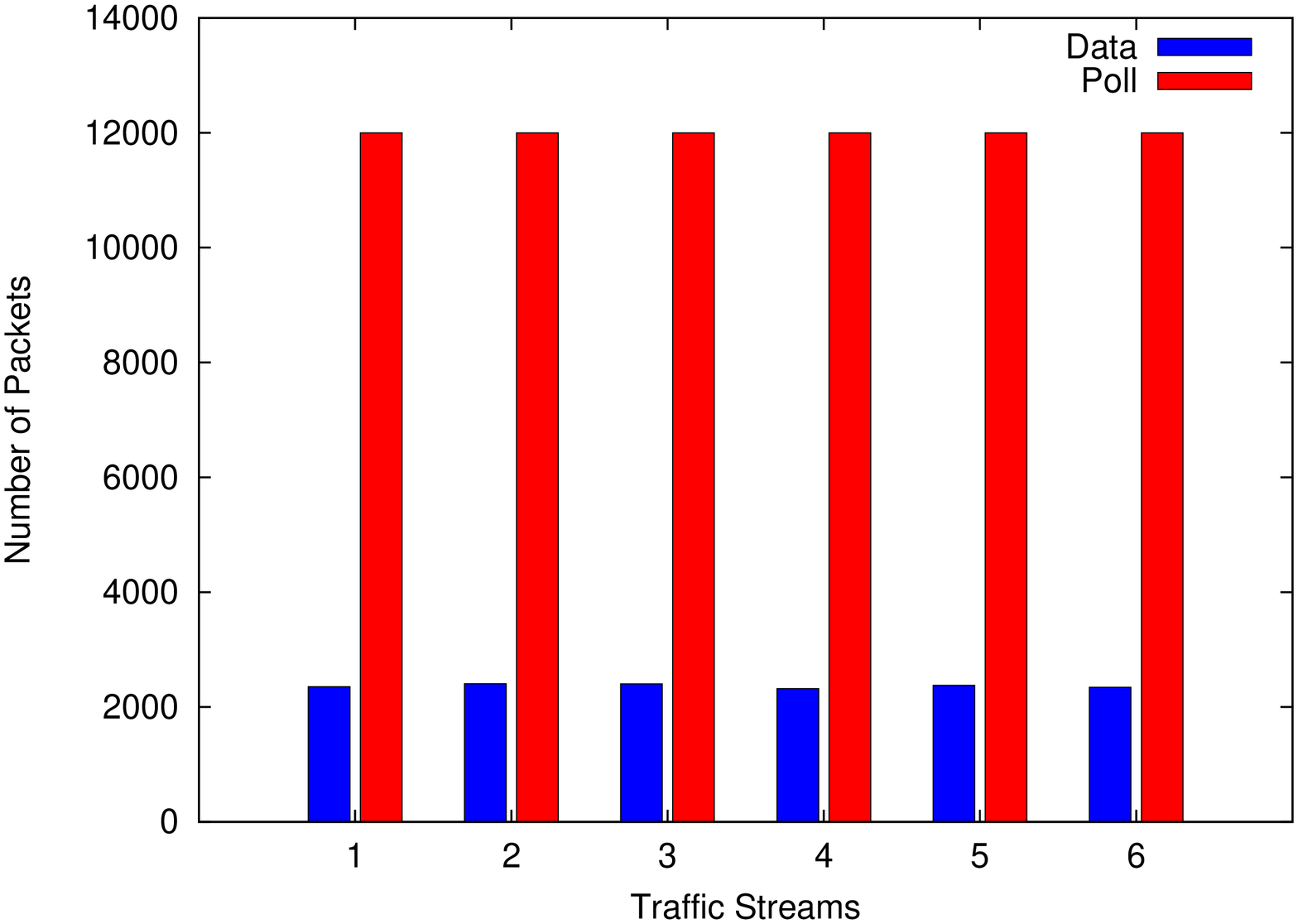}
	}
	\caption{Polls vs Packets of Enhanced EDD as a Function of The Number of Stations.}
	\label{fig:pollsvspktEDD}
\end{figure}

\begin{figure}[!t]
	\centering
	\subfigure[Formula1 Video]{
		\label{fig:pollsvspkt1Fpoll}
		\includegraphics[angle=-90,scale=0.22]{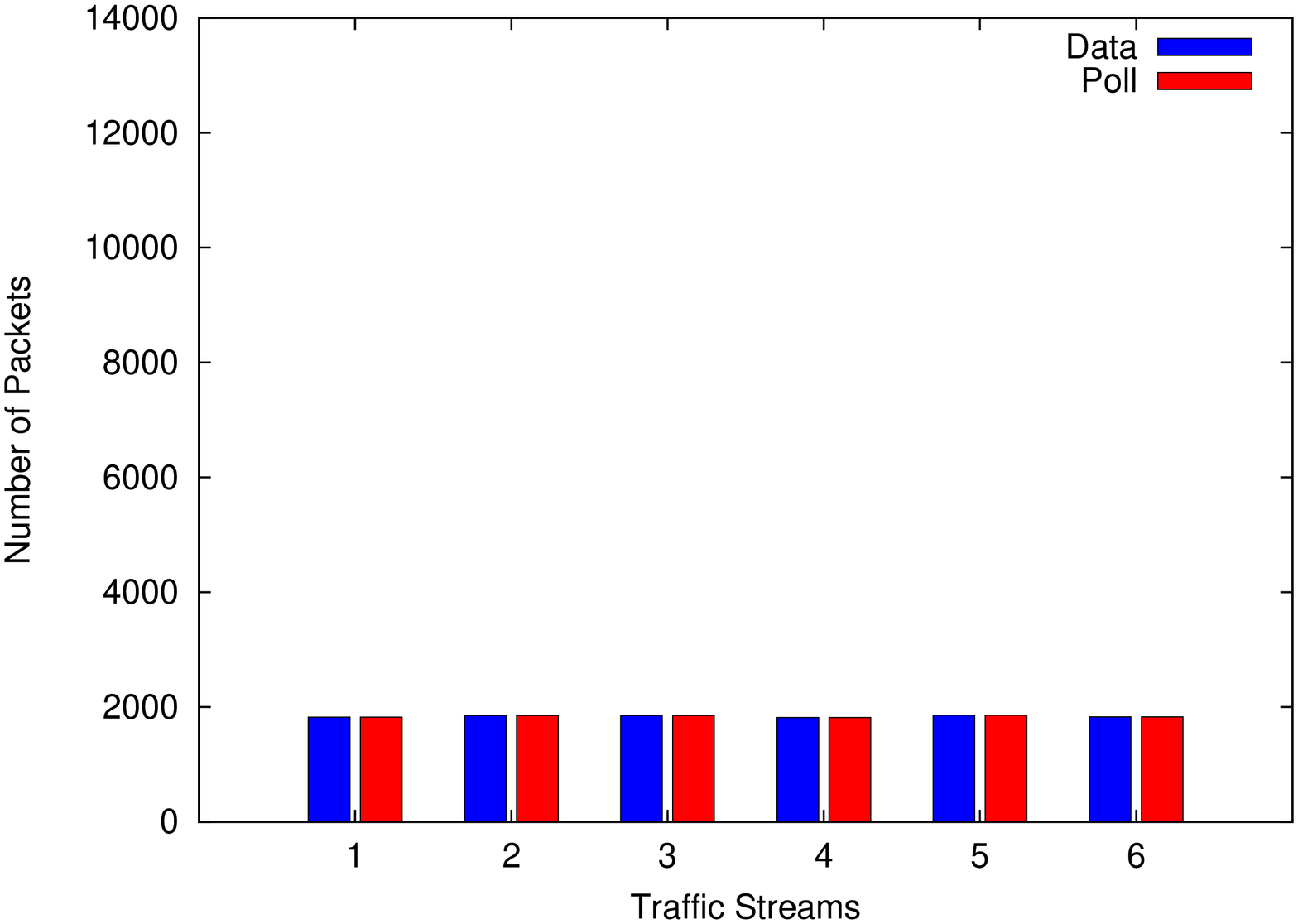}
	}
	\subfigure[Soccer Video]{
		\label{fig:pollsvspkt2Fpoll}
		\includegraphics[angle=-90,scale=0.22]{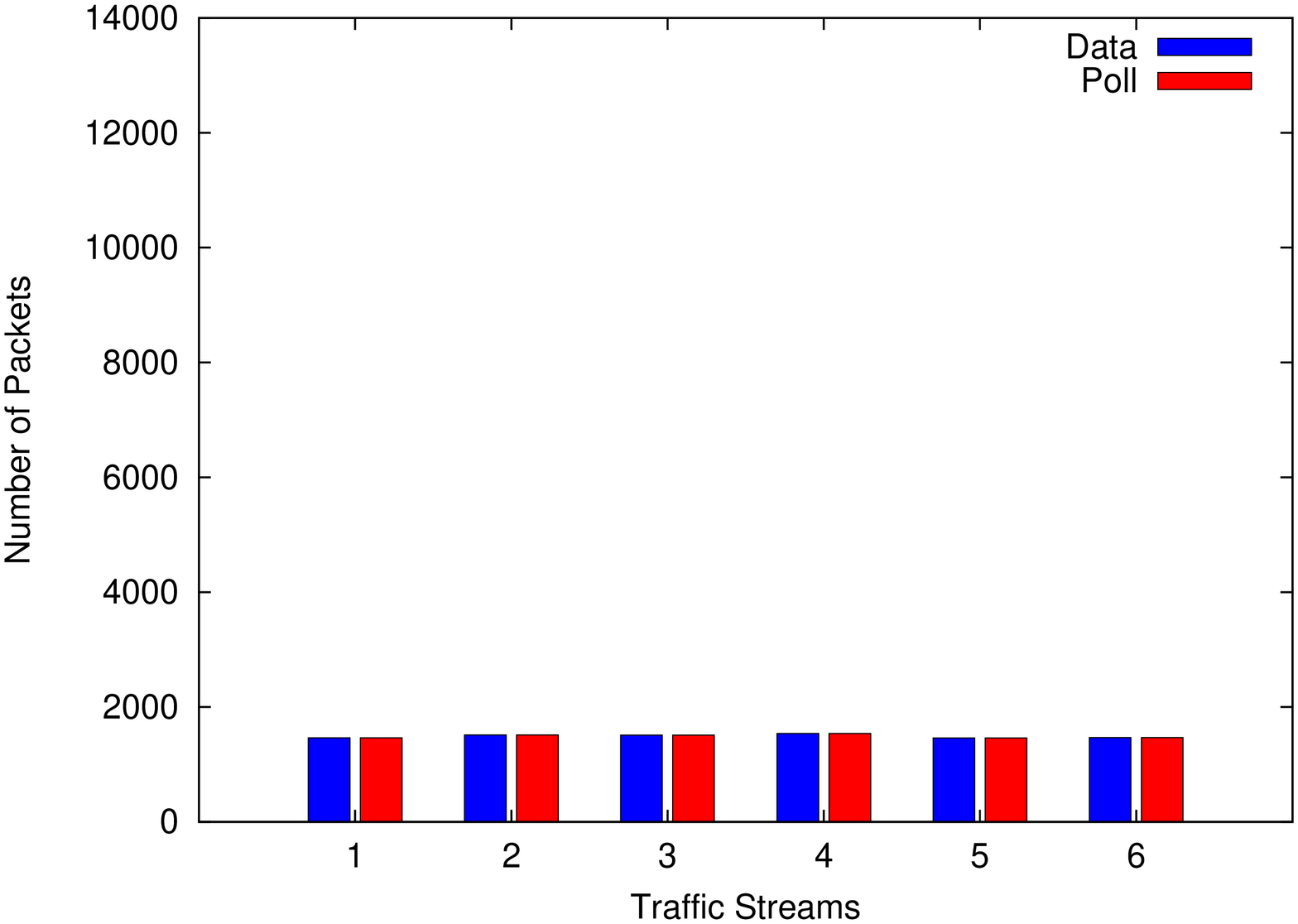}
	}
	\subfigure[Mr Bean Video]{
		\label{fig:pollsvspkt3Fpoll}
		\includegraphics[angle=-90,scale=0.22]{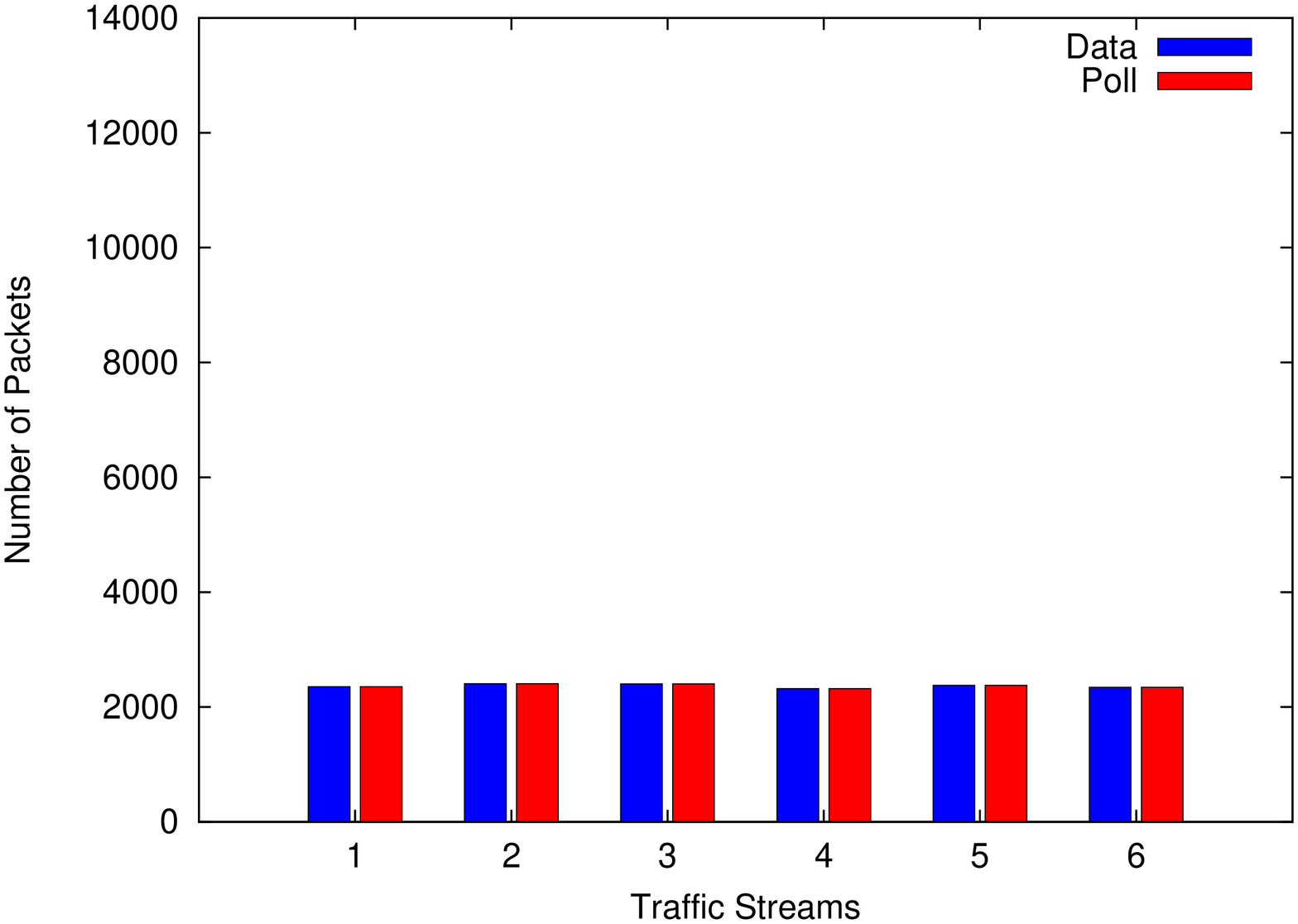}
	}
	\caption{Polls vs Packets of F-Poll as a Function of The Number of Stations.}
	\label{fig:pollsvspktFPoll}
\end{figure}

To further show the effect of the over-polling issue on the channel utilization we demonstrate the number of data packets sent versus the number of poll frames granted to the stations. We have chosen the case of 6 uplink traffics transmitting to the \gls{QAP} of the tested videos.
Figures~\ref{fig:pollsvspkt1Ref}, (b) and (c) as well as \ref{fig:pollsvspkt1edd}, (b) and (c)
show the high number of poll frames versus number of data packets using the reference \gls{HCCA} design and Enhanced \gls{EDD} for the examined video sequences. The massive increase in the number of polls against the actual need is  discussed in
Figures~\ref{fig:pollsvspkt1Fpoll}, (b) and (c). In contrary, Figure~\ref{SIassign} shows the effectiveness of the \gls{F-Poll} scheme in accurately poll stations using the feedback information about the next frame arrival time. 
\subsubsection{Throughput Analysis}
\label{subsubsec:5_thrp}
\begin{figure}[!t]
	\centering
	\subfigure[Formula1 Video]{
		\label{fig:thrp1}
		\includegraphics[angle=-90,scale=0.22]{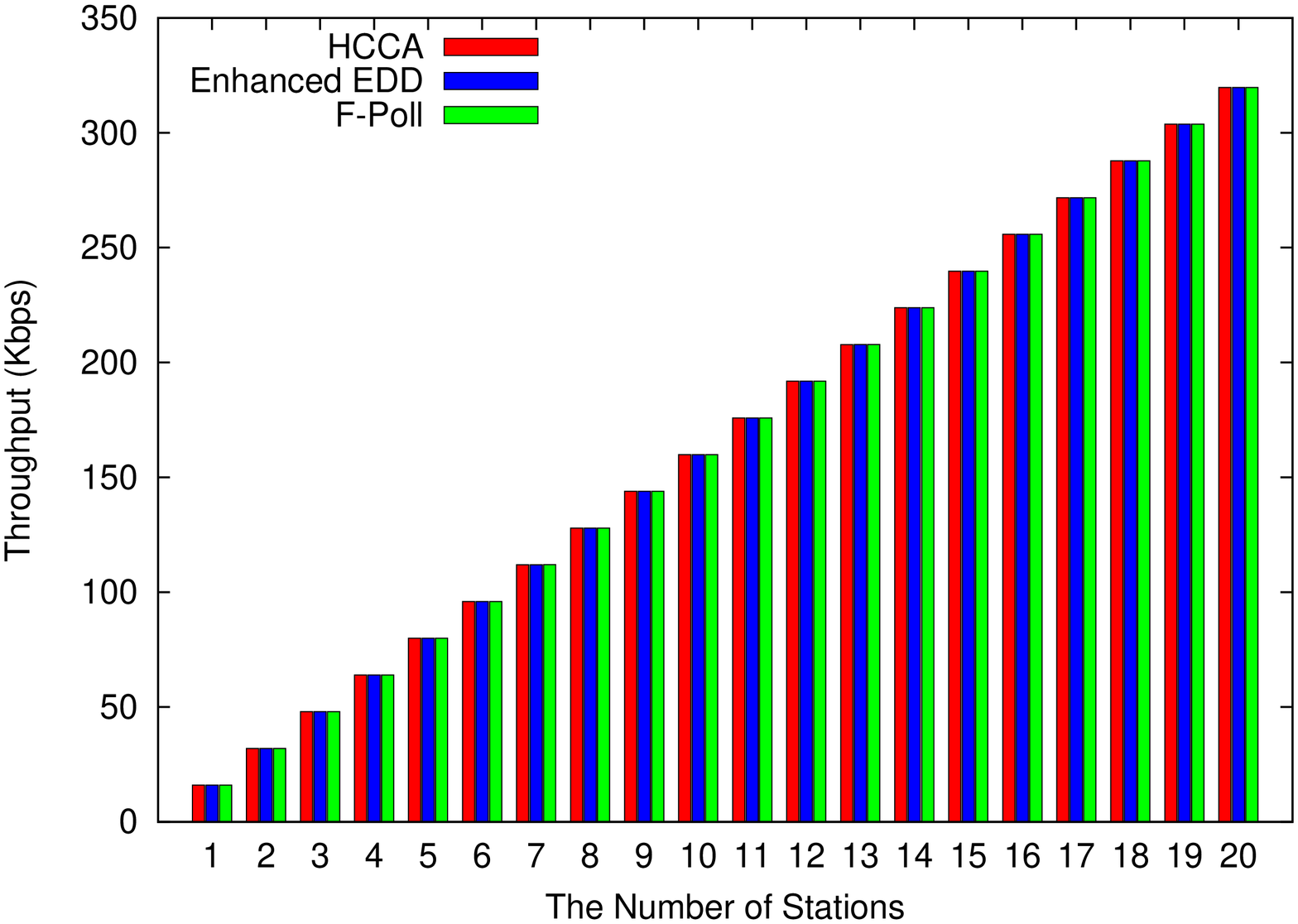}
	}
	\subfigure[Soccer Video]{
		\label{fig:thrp2}
		\includegraphics[angle=-90,scale=0.22]{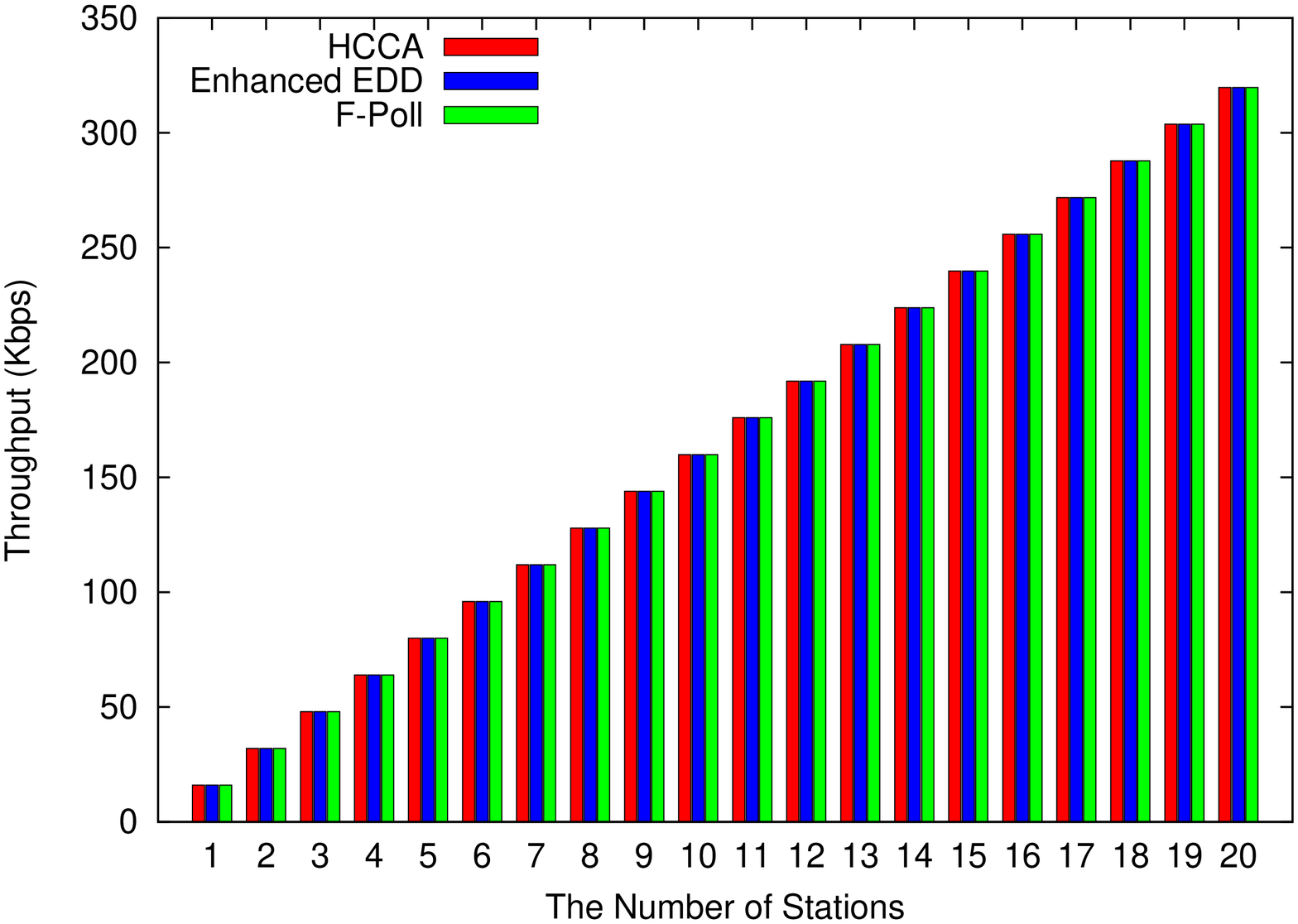}
	}
	\subfigure[Mr Bean Video]{
		\label{fig:thrp3}
		\includegraphics[angle=-90,scale=0.22]{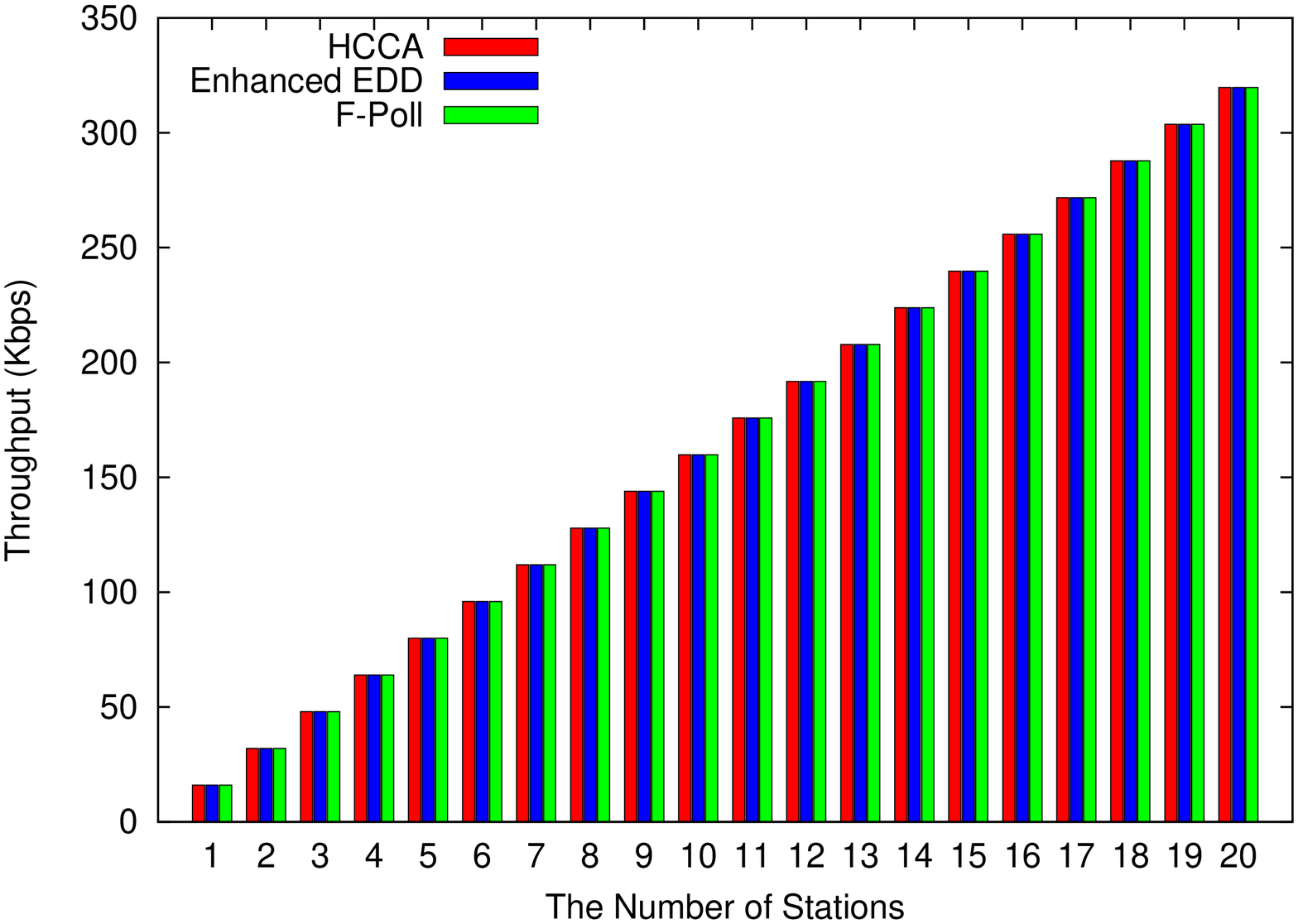}
	}
	\caption{Aggregate Throughput of as a Function of the Number of Stations.}
	\label{fig:thrp}
\end{figure}

We have investigated the aggregate throughput of the examined schemes as a function of the number of stations to verify that our scheme is efficient in supporting \gls{QoS} for \gls{VBR} traffics and maintaining the utilization of the wireless channel. Aggregate throughput is calculated in Equation \eqref{eq:throughput}.
\begin{eqnarray}
\label{eq:throughput}
AggregateThrp  = \frac{1}{T} \sum_{i=1}^{N} ( Size_{i}) ,
\end{eqnarray}
where $Size_{i}$ is the total received packet size at the \gls{QAP}, $T$ is the simulation time and $N$ is the total number of the received packets at \gls{QAP} during the simulation time.
Figures \ref{fig:thrp1}, (b) and (c) depict the aggregate throughput as a function of number of \gls{QSTA}. Since the \gls{F-Poll} scheme utilizes the $QS$ field defined in the standard \gls{MAC} header format to send information about the arrival time of the next frame, no extra overhead is added to the network. As a result, the  \gls{F-Poll} scheme minimized the packet delay while maintaining the throughput similar to that gained using reference \gls{HCCA} design.
\section{Conclusion}
A new polling mechanism has been proposed in this paper to support pre-recorded \gls{VBR} video stream transmission in IEEE 802.11e
\gls{HCCA} networks. \gls{F-Poll} mechanism is a feedback-based mechanism in which the station sends information with each packet sent about the next arrival time of the next frame. Based on this, in each \gls{SI} period, the \gls{QAP} will selectively poll stations that are ready to transmit in order to reduce the poll overhead and thus minimized the delay in the system. Simulation results reveal the efficiency of the \gls{F-Poll} mechanism over both \gls{HCCA} and Enhanced \gls{EDD} polling mechanisms in minimizing the data packet delay and conserve the channel bandwidth by remarkably reduce the poll overhead in the system.
\label{sec:conclusion}





\section*{Acknowledgement}
This work has been supported by the Malaysian Ministry of Education under the Fundamental Research Grant Scheme FRGS/1/11/SG/UPM/01/1.



%
%
\end{document}